\documentclass{iopjournal}

\usepackage[backend=biber,style=numeric,sorting=none]{biblatex}
\usepackage{float} 
\usepackage{enumitem}
\usepackage{booktabs}
\usepackage{siunitx}
\usepackage{subcaption}

\begin{document}

\articletype{Paper} 

\title{Additively manufactured Shape Memory Alloy Hybrid Composites with a polymer matrix featuring a re-entrant honeycomb structure }

\author{Manuel Kunzler$^1$\orcid{0009-0000-7545-8997}, Sascha Bruk$^2$\orcid{0000-0003-4575-5151}, Max Kaiser$^{1,*}$\orcid{0009-0002-5769-232X} and Martin Gurka$^{1,*}$\orcid{0000-0002-6359-5756}}

\affil{$^1$Department of Smart Composites \& Nondestructive Testing, Leibniz Institut für Verbundwerkstoffe GmbH, Kaiserslautern, Germany}

\affil{$^2$Department of Tailored Lightweight Composites, Leibniz-Institut für Polymerforschung Dresden e. V., Dresden, Germany}

\affil{$^*$Author to whom any correspondence should be addressed.}

\email{martin.gurka@leibniz-ivw.de}


\begin{abstract}
Stereolithography (SLA) and Tailored Fiber Placement (TFP) were combined to fabricate shape memory alloy hybrid composites (SMAHC) featuring a three-layer structure and exhibiting out of plane bending deformation when activated, in a fully integrated, additive manufacturing process. SMA wires as active elements were attached to a textile reinforcement layer, which then was embedded within a UV-curable polymer matrix and combined with a geometrically tailored toplayer, featuring the re-entrant honeycomb architecture. Exploiting the design freedom of SLA, the overall mechanical response of the SMAHC can be systematically adjusted, enabling controlled out-of-plane bending during thermal activation. Two different SMA integration strategies—manual embedding and automated TFP were investigated to assess their influence on actuation behavior, reproducibility, and deformation behaviour. A total of eight geometric configurations were manufactured and experimentally characterized using synchronized optical measurements. The results demonstrate that the combination of SLA-based fabrication and textile-mediated SMA integration enables precise control over the actuation response, while the use of re-entrant honeycomb structures provides an effective approach to tailor stiffness and deformation characteristics. In particular, the automated TFP integration yields improved reproducibility and more symmetric deformation behavior compared to manual fabrication. The presented approach establishes a fully additive manufacturing route for SMAHCs, enabling the realization of structurally integrated, morphing composite systems with programmable mechanical properties.
\end{abstract}

\section{Introduction}
\begin{figure}
 \centering
        \includegraphics[width=0.5\textwidth]{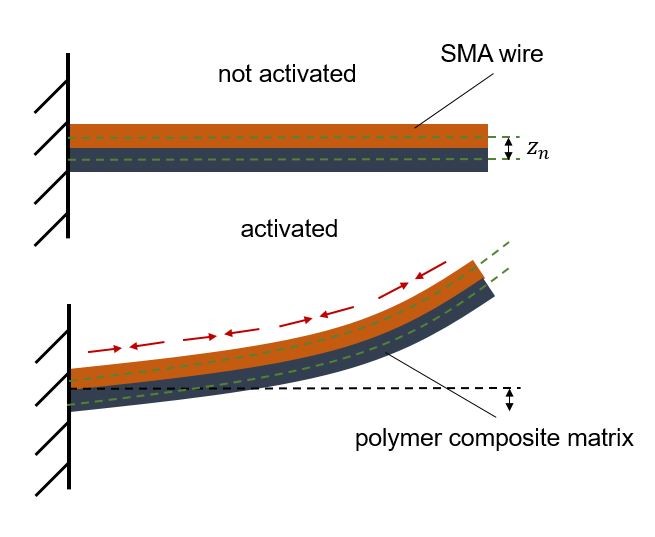}
 \caption{Schematic representation of an unactivated and an activated SMAHC as a bending actuator}
 \label{fig1}
\end{figure}

Shape memory alloys (SMAs) are intermetallic materials capable of a reversible solid–solid phase transformation, enabling a recoverable inelastic strain of approximately 1–8\% under suitable thermal and mechanical loading. Their functional response can be described by the shape memory effect, where pseudoplastic deformation is reversed upon heating and, above a critical temperature, superelasticity, which allows large, reversible strains. These characteristics make SMAs promising candidates for applications requiring precise actuation and adaptive structural behavior \cite{lagoudas_d_c_shape_2008}. Incorporating SMAs into composite materials has garnered significant attention, particularly in developing morphing and adaptive lightweight structures. The foundational idea of Shape Memory Alloy Hybrid Composites (SMAHCs), which merges SMA wires with polymer composites at the material level, emerged in the early 1990s \cite{rogers_development_1988,rogers_shape_1988,}. When the SMA filament is placed out off the neutral axis within the SMAHC structure, it enables active shape manipulation in an out of plane bending deformation \cite{breuer_chapter_2026}.  
In the unactivated state, the composite structure remains undeformed. Activation by an external heat source or Joule heating of the SMA induces a contraction, generating a strain mismatch with the polymer matrix and resulting in an out of plane bending of the composite (Fig. \ref{fig1}). 
Due to their stimulus-responsive shape adaptation, SMAHCs enable advanced functionalities in aerospace \cite{nissle_adaptive_2018,hartl_aerospace_2007,jodin_optimized_2018}, automotive \cite{huebler_active_2015,concilio_shape_2021}, or civil engineering applications \cite{concilio_shape_2021}.
In the literature, there are various approaches to distinguishing the integration scale of SMA in composite materials. In \cite{rodino_design_2024} the authors differentiate between “fully embedded”, in which the SMAs are completely encapsulated in the composite structure and “hybrid”, where the active SMA elements are outside the morphing structure. However, there is need for a more detailed distinction. In \cite{breuer_chapter_2026}, the hybrid case is further refined. The authors distinguish three cases:
\begin{itemize}
  \item \textbf{Microscopic scale:} SMA wires are directly integrated into the composite structure (e.g. \cite{balta_embedded_2001, wu_woven_2013, han_shape_2016,bodkhe_3d_2020}).
  \item \textbf{Mesoscopic scale:} SMA wires are attached to the surface of the composite structure (e.g. \cite{huebler_fiber-reinforced_2016, walgren_development_2018}.)
  \item \textbf{Macroscopic scale:} SMA wires are connected to the surface of the composite structure via additional load bearing elements (e.g. \cite{kunzler_design_2022, kaiser_airfoil_2022,wan_a_hamid_design_2019, jodin_optimized_2018,barbarino_wing_2009, poulin_development_2017}).
\end{itemize}
Based on the findings in \cite{kunzler_design_2022} and \cite{kaiser_airfoil_2022} it is revealed that for larger structures, a differential (macro-scale) approach proves problematic. Undulation and local buckling of the SMA wires degrade reliability and repeatability, length tolerances of the SMA wires cause uneven load sharing and inconsistent response and a multi-component concept increases manufacturing complexity. These drawbacks make macro-level differential designs unattractive, whereas material-level SMA integration seems to be a more promising approach. In the early stages, the fabrication of the SMAHCs was primarily based on conventional composite manufacturing techniques such as prepreg-autoclave processing \cite{pattar_review_2019}. 
More recently, however, an increasing number of studies have investigated additive manufacturing (AM) as a promising route to overcome design limitations and to enable complex architectures. Among these, fused deposition modeling (FDM), direct ink writing (DIW) or polyjet printing have been most frequently employed. The different approaches exhibit inherent challenges. In the case of FDM, the high processing temperatures often trigger premature activation of the SMA during fabrication. To mitigate this issue, many researchers have adopted semi-integrative approaches, in which polymer inlets are first printed and the SMA elements are subsequently inserted post-process \cite{wang_experimental_2023, rodino_design_2024,acevedo-velazquez_manufacturing_2024,onder_development_2024, rajagopalan_bi-state_2022}. 
DIW, by contrast, typically relies on solvent-based inks. In \cite{bodkhe_3d_2020} the authors presented a solvent-based direct inc writing 3D printing process enabling complex geometries. Another example of a solvent-assisted manufacturing approach is presented in \cite{islam_3d_2024}, where the authors demonstrate a “Pause-and-Go” FDM-based fabrication method in which Nitinol wires are manually embedded into a 3D-printed thermoplastic polyurethane substrate softened with a non-polar solvent, resulting in defect-free SMA integration. However, solvent-based manufacturing techniques face several challenges, including the handling of volatile and often hazardous chemicals as well as considerable dimensional shrinkage, which collectively limit their practical applicability.
Another issue is the load transfer. In \cite{ashir_development_2020} it was proposed to integrate SMA wires into glass fiber textiles using Tailored Fiber Placement (TFP). In this configuration, load transfer does not rely solely on the direct interface between SMA and polymer matrix but is mediated by the surrounding textile architecture, which significantly reduces the likelihood of interfacial debonding. Against this background, hybrid concepts combining 3D-printed matrix structures with embedded fabrics appear particularly promising, as they combine the design freedom of AM with the mechanical advantages of textile reinforcement. 
The need for more complex, load-adapted structures is further emphasized in \cite{srivastava_-coupling_2022} where the authors focused on optimizing the numerical bending behaviour of SMA fiber-reinforced composites. The presented SMAHC, firstly introduced in \cite{srivastava_design_2018} consists of SMA wires embedded into an E-glass-fiber fabric and silicone rubber. In a multi-objective optimization simulation approach, the authors showed that a complex structure such as re-entrant honeycombs can significantly improve the damping behaviour as well as a better shape control. However the authors couldn't provide a manufacturable prototype of their simulation approach. In this study, we present an SMAHC with such an re-entrant honeycomb structure in its polymer part, that was additively manufactured via stereolithography (SLA) 3D printing, in which SMA wires are integrated within a glass-fiber fabric as an additional reinforcement and load transfer structure. Two reinforcement variants were employed: in the first, the SMA wires were manually integrated into the fabric (referred to as hand-woven, HW), whereas in the second, the wires were positioned on the glass fiber fabric using a TFP process. SLA was chosen to overcome the drawbacks of hot extrusion- or solvent-based approaches, as it enables fabrication at room temperature, the use of flexible photopolymers, and the integration of fiber reinforcements to enhance load transfer and structural compliance.
\section{Methods and Materials}
\label{sec:method and materials}
\subsection{General concept}
\label{sec:general concept}
The design of the SMAHCs investigated in this study is based on the idea of combining SMA wires with an elastic substrate whose mechanical properties can be purposefully tailored by additively manufacturing different honeycomb geometries, to influence its actuation response. To achieve this, a two-layer architecture was selected, consisting of an active SMA layer (actuatorlayer - AL) and a passive toplayer (TL) (see Fig. \ref{fig2}). 

\textit{The actuatorlayer (AL)}: In the AL, the SMA wires are integrated into a glass-fiber fabric, forming a glass-fiber–SMA preform that provides mechanical anchorage for the wires. This preform is subsequently embedded into the photopolymer matrix during the SLA printing process. As a result, the actuation forces generated by the SMA wires are distributed more evenly into the surrounding composite structure through the additional glass-fiber reinforcement.
\textit{The toplayer (TL)}: The TL represents the flexible substrate that ensures that the SMA wire is positioned outside the neutral axis of the entire SMAHC. As the TL is manufactured via SLA printing it can be engineered in a way to modify the overall stiffness and damping characteristics of the SMAHC, ensuring  a specifically controlled deformation during actuation. Another key aspect is the possibility to introduce any stiffness distribution in the TL, therefore being able to tailor the bending moment distribution and, consequently, the curvature and deflection behavior of the entire SMAHC under load. 
\begin{figure}[H]
 \centering
        \includegraphics[width=0.95\textwidth]{layers_of_SMAHC.png}
 \caption{Schematic representation of the structure of the SMAHC, with an actuator- and a toplayer. Depicted are also the neutral axes of the actuatorlayer, the toplayer and the resulting neutral axis of the entire SMAHC}
 \label{fig2}
\end{figure}
To ensure a good comparability of the experimental results, a uniform geometry of the SMAHC was defined, which differs only in the thickness of the re-entrant honey comb cell structure. The re-entrant layer has a length of approximately 84 mm and a width of 66 mm. At the base of the SMAHC, a 90 mm wide solid reinforcement rib is integrated to facilitate secure mounting within the test bench. An additional solid rib is located at the front side, providing designated positions for the attachment of measurement points. Fig. \ref{fig3} shows the corresponding CAD design.
\begin{figure}[H]
 \centering
        \includegraphics[width=0.75\textwidth]{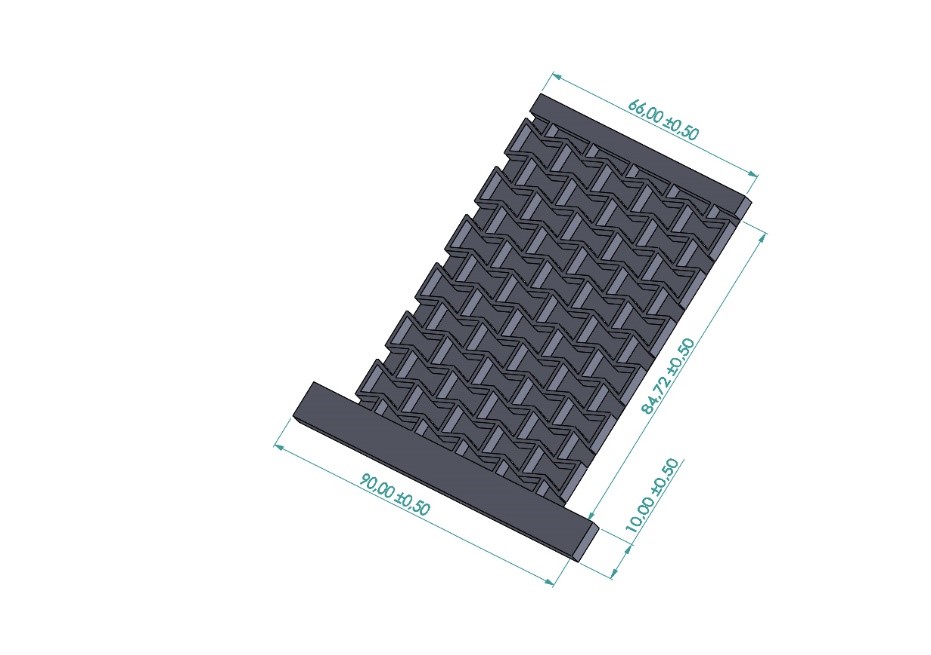}
 \caption{CAD design overview of the SMAHC with its reentrant honeycomb structure and a solid reinforcement rib at both ends (all measures in mm). }
 \label{fig3}
\end{figure}

\begin{figure}[H]
 \centering
        \includegraphics[width=0.75\textwidth]{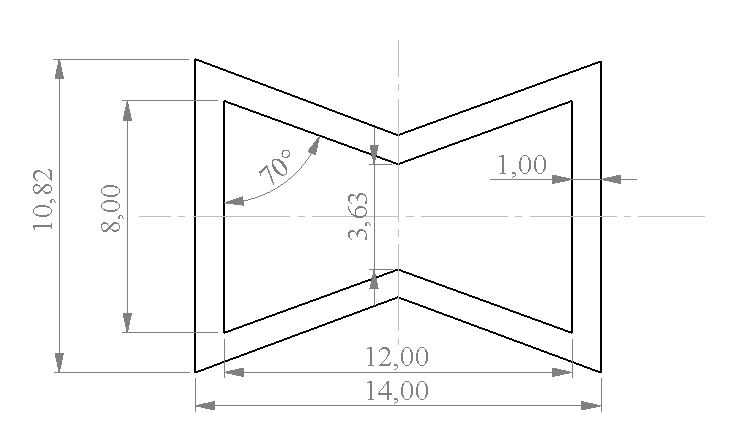}
 \caption{Geometric details of the re-entrant honeycomb unit cell (all measures in mm).}
 \label{fig4}
\end{figure}

The actuatorlayer, consisting of the glass fiber fabric with integrated SMA wires has a constant thickness of approx. 1 mm. To explore the various possibilities of the manufacturing method, three different toplayers with re-entrant honeycomb structure (3 mm, 4 mm and 5 mm) were manufactured in order to vary the resulting position of the neutral axis of the SMAHC. Fig. \ref{fig4} illustrates the key dimensions of the re-entrant honeycomb unit cell, which is integrated as the core component of the toplayer. 
Each specimen is composed of 50 identical unit cells.
The unit cells feature a uniform wall thickness of 1 mm and a re-entrant angle of 70°, defining the auxetic behavior of the structure.
\subsection{Materials}
SmartFlex Nitinol wires from SAES Getters with a diameter of 500 µm were used. Due to thermomechanical stabilization in the manufacturing process, the SMA wires have a defined stress–strain state in the range of 50 MPa to 400 MPa and typical strain values of 1 to 5 \%. The austenite start and finish temperatures of the material are 86 °C and 90 °C, respectively.
Polymer matrix was BASF Ultracur3D EL 4000, which is a UV-curable flexible photopolymer resin with an elongation at break of 172 \%, a hardness of 90 Shore A, a rebound resilience of 30 \%, and a Young’s modulus of 42 MPa. 
For the glass-fiber fabric, unidirectional SAERTEX U-E-640g/m²-1260mm was used. It is composed of a primary 0° layer with a basis weight of 567 g/m² made of E glass and supplementary layers for mechanical stabilization in 45°, 90° and -45° orientation (each approx. 13 g/m² and 35 g/m² respectively), supplemented by a synthetic polyester stitching (12 g/m²), resulting in a total basis weight of 640 g/m².
The reinforcement stitching was performed using a Kevlar/aramid sewing thread (size 75, Extremtextil, Germany) composed of 100 \% para-aramid fibers. The thread exhibits a fineness of 50/2 Nm (Tex 40) and is designed for needle sizes NM 80–100. It withstands continuous operating temperatures up to 300 °C and short-term exposure up to 425 °C.
For the glass-fiber-SMA preform production using TFP, glass fiber roving SE 1200 from OWENS CORNING with 1200 tex made of E-CR glass was used. A woven textile made of  glass fiber with 104 g/mm² served as base material. Polyester thread with 10 tex (SERAFIL 1000 200/2) was used as upper and lower stitching thread in the embroidey machine (TAJIMA TCWM-101).

\subsection{Manufacturing process}
\label{sec: Manufacturing process}
The manufacturing process of the SMAHC consited of five steps.
Pre-stretched SMA wires were positioned within a glass-fiber fabric, either manually or by means of the TFP process.
The resulting glass fiber–SMA preform which then defines the alignment and anchorage geometry of the wires was subsequently placed in the SLA printer.
During printing, the liquid photopolymer resin impregnated the textile structure and was cured layer by layer under ultraviolet light, resulting in a fully embedded composite with homogeneous bonding between the matrix and the reinforcement.
The five steps of the manufacturing process in more detail are as follows: \\
(1) Preparation of the SMA wires was done by tempering at 100°C for 60 minutes to eliminate any de-twinning effects from storage or previous handling. To be able to call up the one-way shape memory effect, it was then pre-stretched by 4\% using a universal testing machine (UTM 1445, ZwickRoell, Germany). This causes well defined detwinning of the martensitic crystal lattice. Furthermore, the wire surface was cleaned with isopropanol to ensure the best possible adhesion to the matrix material. \\
(2) For the manually integrated specimens, the pre-stretched SMA wire was locally deactivated at the meander-shaped anchorage points in order to prevent unwanted actuation in these regions. 
Electric current with 3\,A was applied to the wire at the selected positions using manually attached electrodes, resulting in localized Joule heating and a structural phase transformation above the austenite finish temperature. 
Temperature was monitored using a thermal camera (TIM 640, Micro-Epsilon, Germany) to ensure sufficient heating. 
For the specimens manufactured using the TFP process, prior deactivation of the meander anchorage points was not feasible due to the automated integration method. 
In these cases, the anchorage geometry itself provided sufficient mechanical constraint to prevent undesired actuation in the stitched areas.\\
(3) Integration of the SMA wires into the glass fiber reinforcement was carried out using two alternative approaches. For manual integration (HW) the SMA wire was inserted manually into the glass fiber fabric as the weft thread, while bundles of glass fibers threads served as warp. Additionally, at the meander-shaped anchoring points, the SMA wire was stitched to the glass-fiber fabric using an extra kevlar yarn to enhance mechanical interlocking. For automated integration via TFP (TFP), firstly glass fiber preforms with dimension of 150 mm x 90 mm were produced by placing one 90°-layer of the SE1200 GF roving with a spacing of 1.2 mm on woven GF base material with a ± 45° orientation. The spacing of the GF rovings was chosen to obtain a sufficiently dense textile layer without gaps that could lead to wire undulation in laminate thickness direction caused by the tension of the stitching thread. The basis weight of the GF preforms resulted to approximately 1104 g/mm². Then the pre-strained SMA wire (without any deactivated areas) was attached to the GF preforms via TFP, using the same stitching thread as for the glass fibers. A very dense embroidery coverage along the wire, with a transverse stitch spacing of 1 mm, was chosen to fixate the wire in the designed meandering path. At some turning points of the meandering pattern the wire could not always be caught by the upper stitching thread. In these areas the wire was additionally covered with Kevlar-thread by hand.\\

  \begin{figure}[H]
 \centering
        \includegraphics[width=0.9\textwidth]{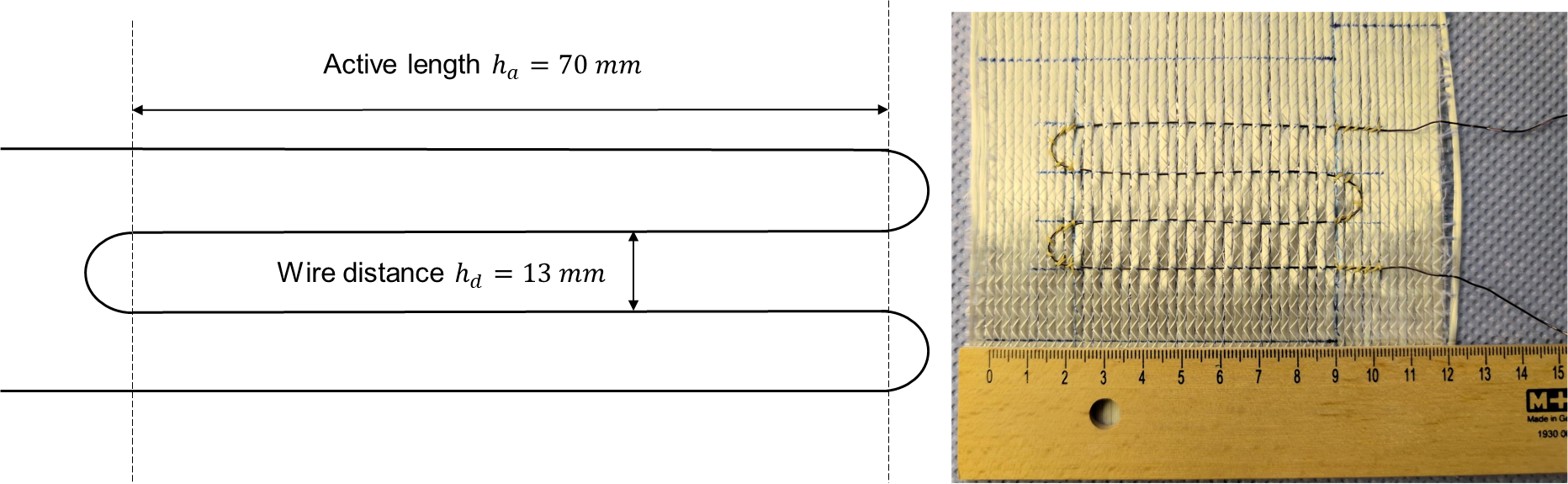}
        \includegraphics[width=0.9\textwidth]{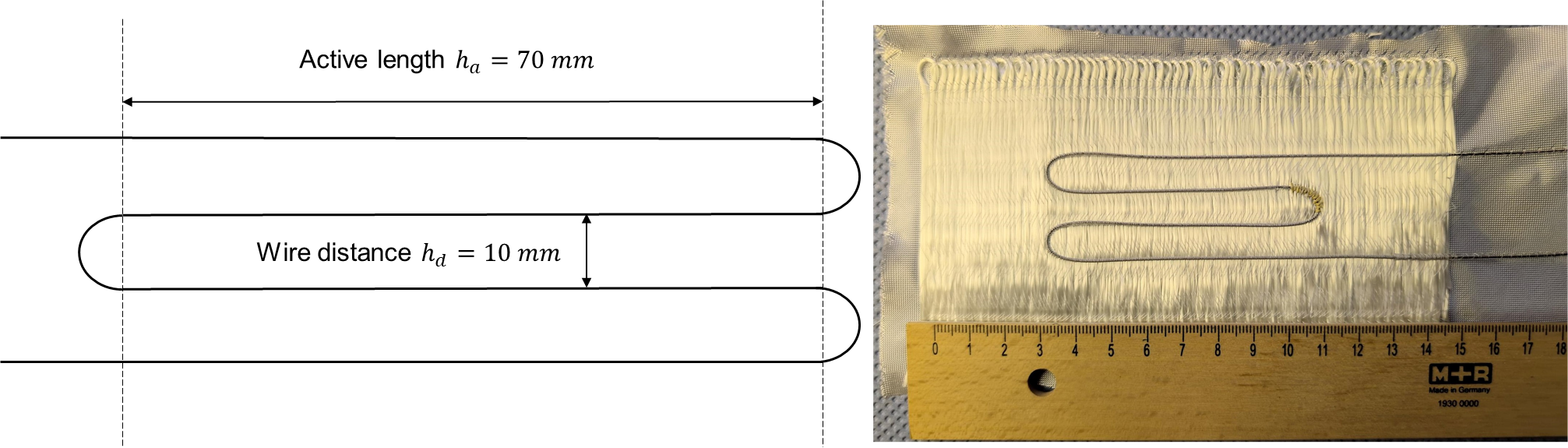}
 \caption{Glass fiber fabric with integrated SMA wire. (top) manually integration (bottom) integration by tailored fiber placement process.}
 \label{fig5}
\end{figure}
  
\begin{figure}[H]
 \centering
        \includegraphics[width=0.9\textwidth]{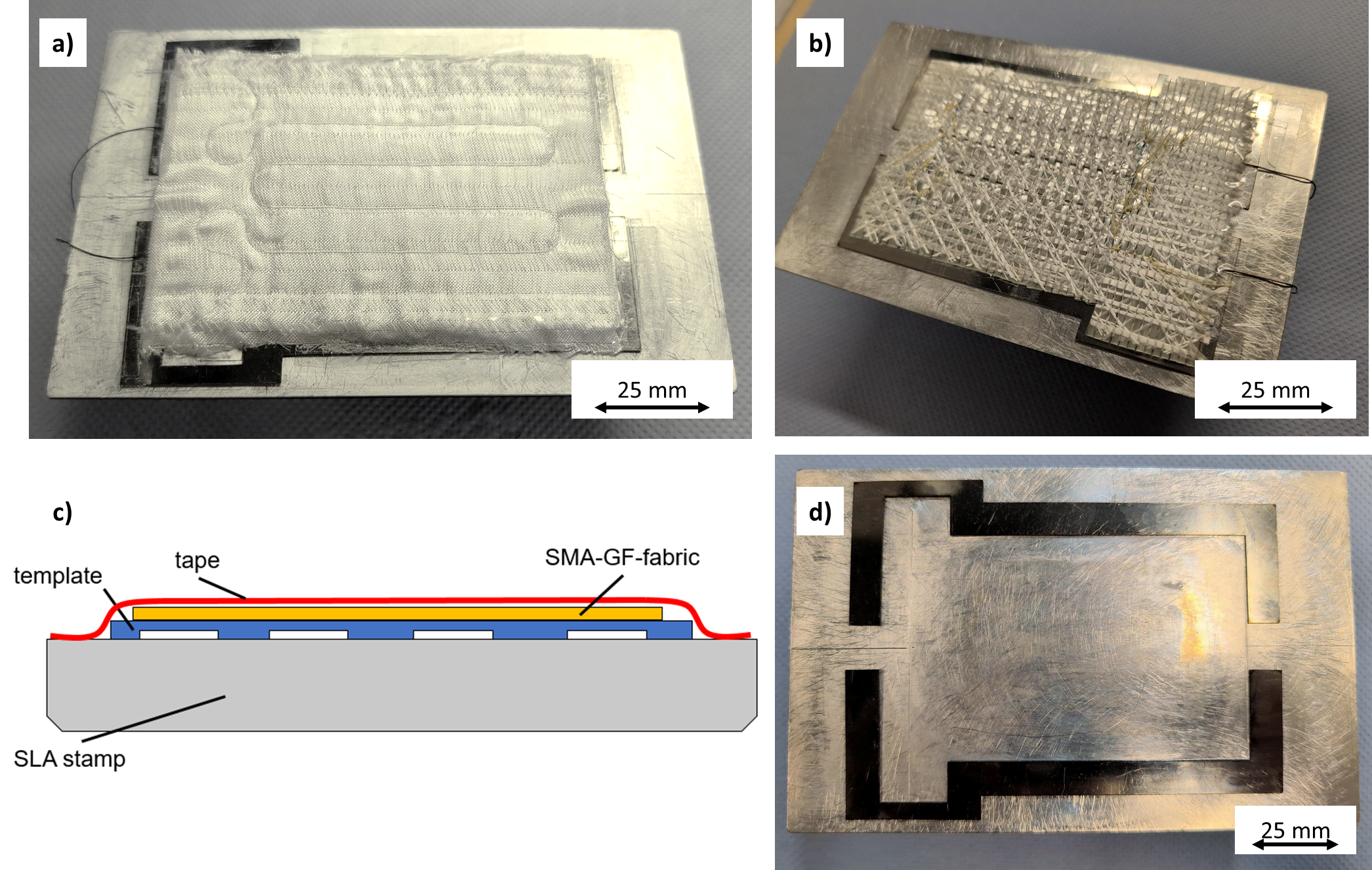}
 \caption{Buildplates}
 \label{fig77}
\end{figure}
(4) Before the printing process, both the manually produced and the TFP-manufactured glass-fiber–SMA fabrics were cut to a size of 120 mm × 76 mm to fit the build platform and to ensure a consistent specimen geometry. The additive manufacturing process then was carried out using a stereolithography (SLA) printer (M3, Anycubic Photon, China) equipped with a 4K monochrome LCD exposure system. Since the original printer configuration did not provide an adjustable build-platform offset, the system was mechanically modified to allow precise control of the printing gap. For this purpose, the optical limit switch of the printer’s z-axis was equipped with a fine-pitch adjustment screw, enabling the build-platform offset to be varied with 0.01 mm precision. This modification allowed a repeatable precise positioning of the glass-fiber–SMA preform on the build platform, ensuring the correct resin layer thickness above the textile structure during impregnation. To verify and reproduce the set offset for each specimen, the platform position was continuously monitored using a Panasonic HG-C1050 laser distance sensor. This setup ensured that all specimens were fabricated with identical layer gaps and thus a uniform overall thickness. After the offset adjustment was completed, two metallic washers were positioned on the build platform at predefined locations that were marked to ensure repeatable placement for each specimen. The washers were fixed in place using standard adhesive tape. Their function was to improve the impregnation behavior of the resin by allowing trapped air to escape through two narrow openings while the photopolymer resin infiltrated the glass-fiber–SMA preform. Once the washers were installed, the glass-fiber–SMA preforms were placed on top of them and also fixed with adhesive tape to prevent movement during the printing process. The printing stamp (build plate) was then mounted into the printer.

Before starting the actual printing routine, the printer software allows manual lowering of the build plate into the resin bath. This feature was used to immerse the mounted preforms in the liquid resin for an impregnation time of 120 seconds, ensuring complete wetting of the glass-fiber–SMA structure. After the impregnation phase, the SLA printing process was initiated. At the beginning of the build, the printer generated a series of burn-in layers, which in standard SLA processes serve as an initial penetration and anchoring stage to ensure reliable adhesion of the resin to the stamp surface. In the present manufacturing approach, this burn-in phase was deliberately repurposed to print the actuatorlayer, since the high exposure energy and improved wetting during these initial layers promote reliable embedding and fixation of the SMA wires. Once the burn-in sequence was completed, the regular printing process continued. Both the burn-in layers and all subsequent layers were produced using a constant nominal layer thickness of 0.05 mm, as defined in the slicing software. This value represents the process parameter provided to the printer and was not measured directly on the printed parts, but is assumed to be accurately maintained by the machine. Due to the increased overall thickness in the region containing the glass-fiber–SMA preforms (referred to as the actuatorlayer), light transmission through the fibers led to reduced curing intensity in this section. To compensate for this effect and to ensure sufficient polymerization throughout the impregnated region, the initial burn-in layer was exposed for an extended curing time of 60 seconds. The burn-in phase consisted of one burn-in layer followed by four transition layers, which gradually reduced the exposure time to the value used for the regular printing layers. After these transition layers, all subsequent layers were cured with an exposure time of 4 seconds per layer. This exposure strategy was chosen to allow for uniform curing and complete impregnation of the glass-fiber–SMA structure while avoiding overcuring of the surrounding polymer matrix. The total printing time ranged between 45 and 60 minutes, depending on the number of layers and the final thickness of the SMAHC structure. Upon completion of the print, the specimen was left to drain for approximately 45 minutes to allow excess resin to flow off before proceeding to the post-processing stage.
(5) Post curing of the specimen was performed after the final SMAHC layer was printed. The stamp with the attached component was removed and the specimen was subjected to a three-step isopropanol washing process to remove excess resin. The SMAHC then was post-cured in an oven at 40°C for approximately 30 minutes. Once completely dry, it underwent an additional UV post-curing treatment for about 10 minutes. This sequential process ensured that the component achieved the material properties specified in the data sheet.

\par To ensure a systematic comparison between the two integration approaches, specimens were 
manufactured across three toplayer thickness levels (3.0, 4.0, and 5.0\,mm) using both HW and 
TFP architectures. 
Table~\ref{tab:samples} provides an overview of all SMAHC specimens 
investigated in this study, including their sample IDs, toplayer thicknesses, and the 
corresponding integration method. This overview also establishes the nomenclature and 
traceability used throughout the subsequent mechanical and actuation analyses.
\begin{table}[htbp]
\centering
\caption{Overview of the investigated SMAHC specimens including wall thickness and integration method.}
\label{tab:samples}
\begingroup
\footnotesize
\begin{tabular}{l c c}
    \toprule
    \textbf{Sample ID} & \textbf{Toplayer thickness in mm} & \textbf{Integration method} \\
    \midrule
    HC0037 & 3.0 & HW \\
    HC0038 & 3.0 & HW \\
    HC0039 & 3.0 & HW \\
    HC0034 & 4.0 & HW \\
    HC0035 & 4.0 & HW \\
    HC0036 & 4.0 & HW \\
    HC0042 & 5.0 & HW \\
    HC0043 & 5.0 & HW \\
    HC0044 & 5.0 & HW \\
    \midrule
    HC0062 & 3.0 & TFP \\
    HC0063 & 3.0 & TFP \\
    HC0064 & 3.0 & TFP \\
    HC0059 & 4.0 & TFP \\
    HC0060 & 4.0 & TFP \\
    HC0061 & 4.0 & TFP \\
    HC0067 & 5.0 & TFP \\
    HC0068 & 5.0 & TFP \\
    HC0069 & 5.0 & TFP \\
    \bottomrule
\end{tabular}
\endgroup
\end{table}

\section{Experimental Setup and Measurement Procedure}
\label{sec:methods}

To investigate the thermo-mechanical behavior of the SMAHC, a fully automated cyclic heating and cooling experiment was conducted. The aim was to capture the thermo-mechanic response of the composite over multiple actuation cycles to assess both the running in and the stabilization of the actuator response.

\textit{Specimen Preparation} - Distinct marker points were applied on the specimen surface to enable displacement analysis using Digital Image Correlation (DIC). 
A matte black coating with an emissivity of $\varepsilon = 0.97$ was applied to a defined area of the specimen surface to ensure temperature measurement via an infrared camera. 
The specimens were mounted on a rigid test frame ensuring planar alignment and reproducible boundary conditions.

\begin{figure}[H]
 \centering
        \includegraphics[width=0.7\textwidth]{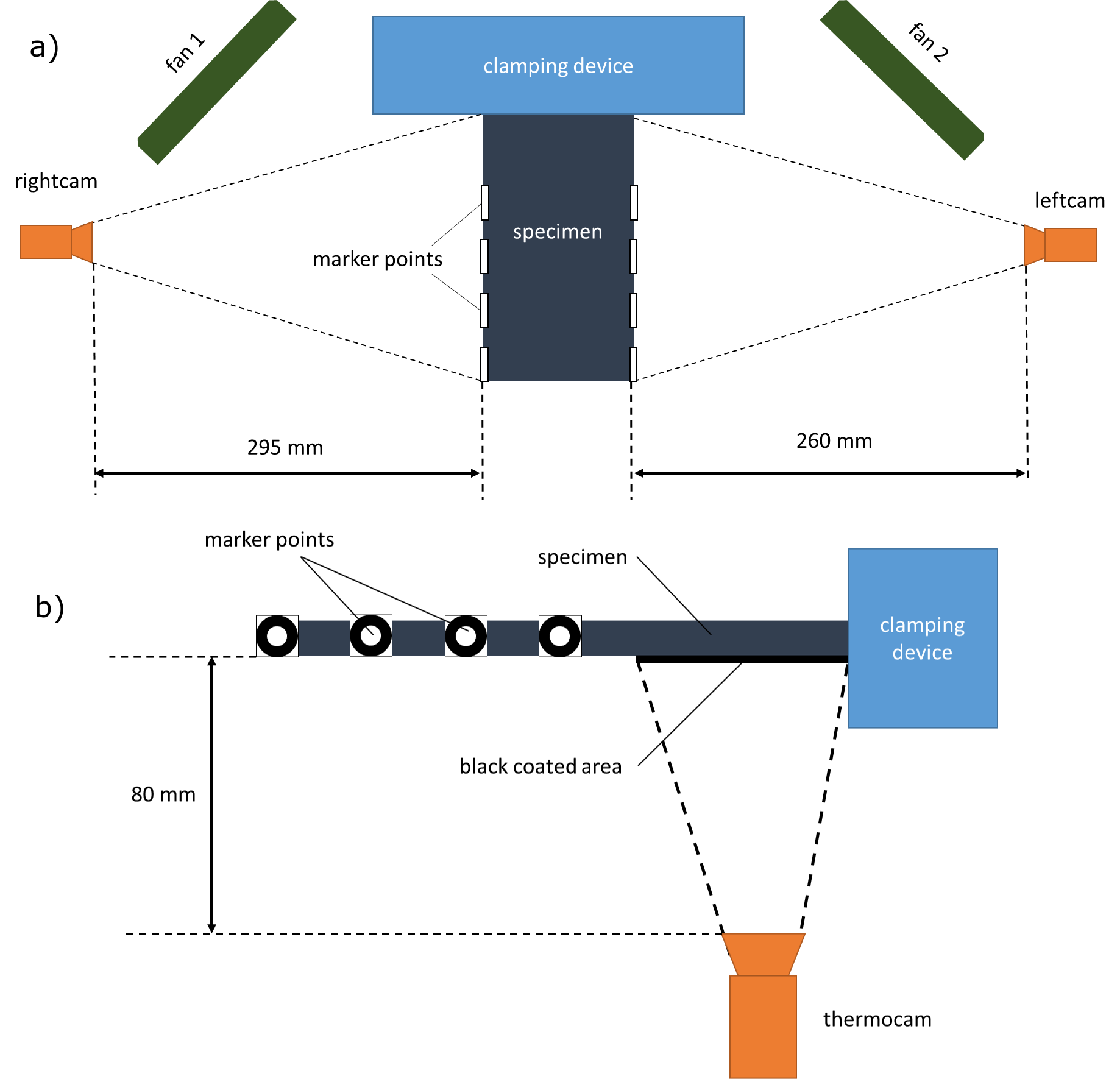}
 \caption{Schematic overview of the experimental setup.
a) Optical configuration for displacement tracking using left and right cameras, including marker point distribution and specimen positioning.
b) Thermal imaging setup showing the black-coated area for emissivity control and the positioning of the infrared camera relative to the clamped specimen.}
 \label{fig7}
\end{figure}

\textit{Optical Measurement System} - Two cameras (IDS U3-3800CP-M-GL R2,  IDS Imaging Development Systems GmbH, Germany) were employed on both sides for simultaneous measurement of both sides of the specimen.
The cameras were synchronized via a shared trigger signal and operated at a frame rate of 10\,Hz with an exposure time of 40\,ms. Diffuse LED illumination was used to enhance contrast and minimize reflections. 
A schematic top and side view of the experimental setup is shown in Fig.~\ref{fig7}.
\subsection{Control and Test Procedure}
\label{sec:control and test procedure}

The experimental sequence for actuation of the SMAHC modules was fully automated using a custom routine written in Python~(3.12.7, 64-bit) 
to control a programmable power supply (HMP~4030, Hameg GmbH, Germany) to drive the embedded SMA wires via controlled Joule heating at a constant current of \SI{4}{\ampere}. 
Each actuation cycle consisted of a heating phase of \SI{13}{\second} and a subsequent cooling phase of \SI{200}{\second}, during which an axial fan was automatically activated to enhance convective heat transfer and accelerate cooling down to ambient temperature. 
After completion of the cooling phase, the next cycle was initiated automatically. 
A total of 50 consecutive cycles were performed for each specimen to capture the evolution and stabilization of the actuation behavior. 




\textit{Temperature Measurement} - The surface temperature on the lower side of the specimen was recorded using a  infrared camera  with a resolution of 640 × 480 px (TIM 640, Micro-Epsilon, Germany), a spectral range of 7.5–13 µm, and a temperature range of 0–250 °C. The camera operated in hot-spot mode, automatically detecting the maximum temperature within the field of view.  Data acquisition and calibration were performed using the TIMConnect software (Rel. 3.21.3113.0, Micro-Epsilon, Germany). The emissivity was set to 0.97 to match the black coated surface of the specimen and verified by reference measurements on the coated surface. The ambient temperature was monitored by a Type K thermocouple, ensuring stable environmental conditions throughout all tests.




\textit{Environmental Conditions} - All tests were conducted in a climate-controlled laboratory at a constant ambient temperature of 20 ± 0.5 °C. The environment was shielded from external light sources; fluorescent ceiling lights provided background illumination. 

\textit{Data Processing} - The recorded video sequences from both cameras were analyzed using digital image correlation software (GOM Correlate Professional 2016, GOM inc., Germany). Marker points applied on the specimen surface served as reference features for the optical tracking. The software uses a facet-based correlation approach, where each facet defines a local subregion for displacement tracking. A facet size of 35\,×\,35\,pixels was chosen, which proved sufficient to ensure robust point tracking throughout all cycles. For calibration of the spatial scale, a 30\,mm ×\,20\,mm adhesive label with a millimeter grid was attached near the active specimen area. Based on this reference pattern, the spatial conversion factor was determined to be 10\,mm\,/\,125.63\, pixels, which remained constant for all videos since the camera configuration was identical for all tests. Separate \texttt{.csv} files were generated for the left and right camera recordings, containing the extracted displacement data for every cycle. 

\section{Results and discussion}
\label{sec:results}
\subsection{General behavior of the actuator response}
Two different glass-fiber SMA preform configurations whith three variants each, all based on a re-entrant honeycomb design but differing in total toplayer thickness, were investigated, corresponding to the \textit{HW} and the \textit{TFP} integration approaches described in Section  \ref{sec: Manufacturing process} and Table \ref{tab:samples}. 

\begin{figure}[H]
    \centering
    
    \begin{subfigure}[b]{0.48\textwidth}
        \centering
        \includegraphics[width=\linewidth]{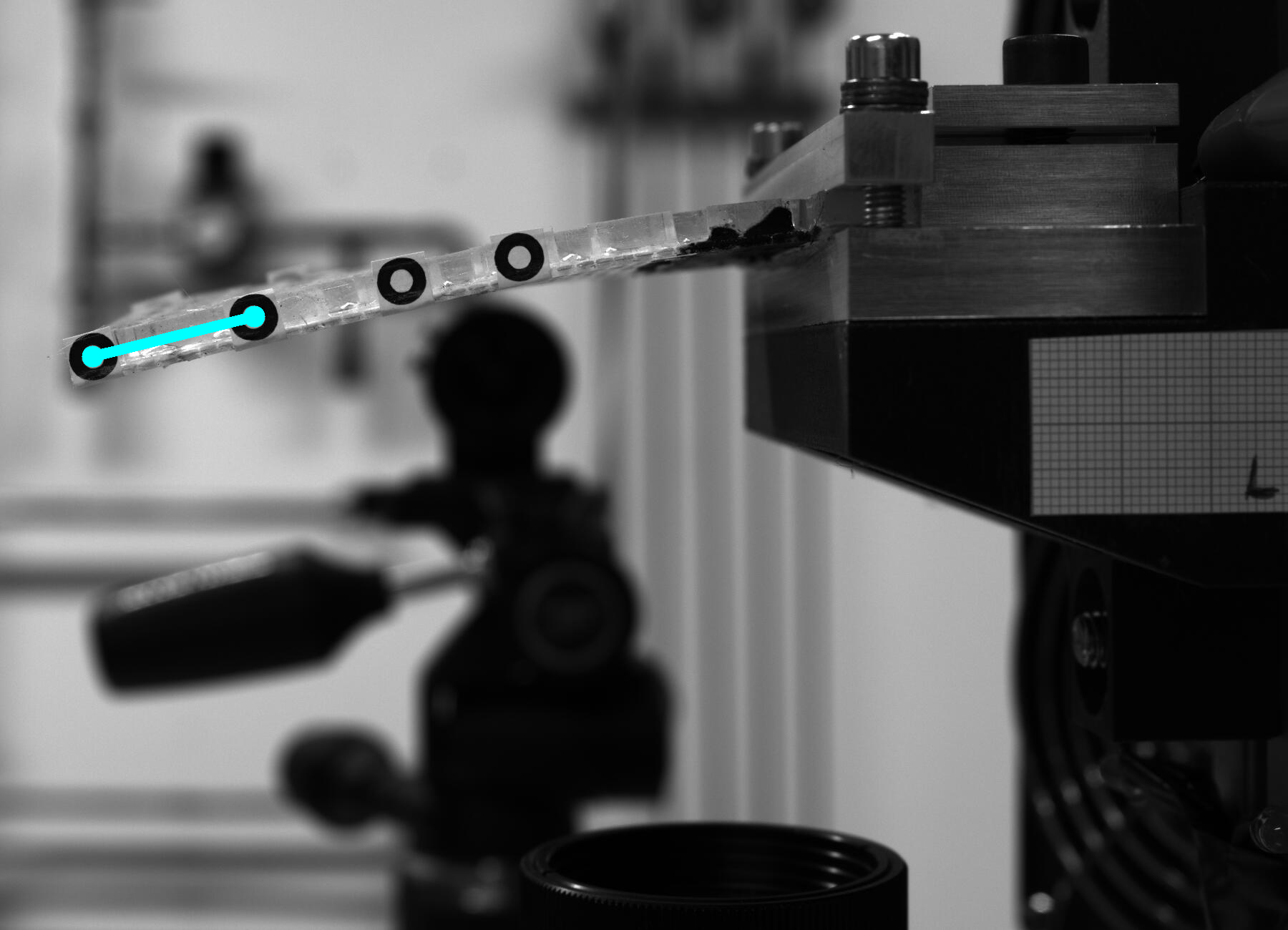}
        \caption{}
        \label{fig:a}
    \end{subfigure}
    \hfill
    \begin{subfigure}[b]{0.48\textwidth}
        \centering
        \includegraphics[width=\linewidth]{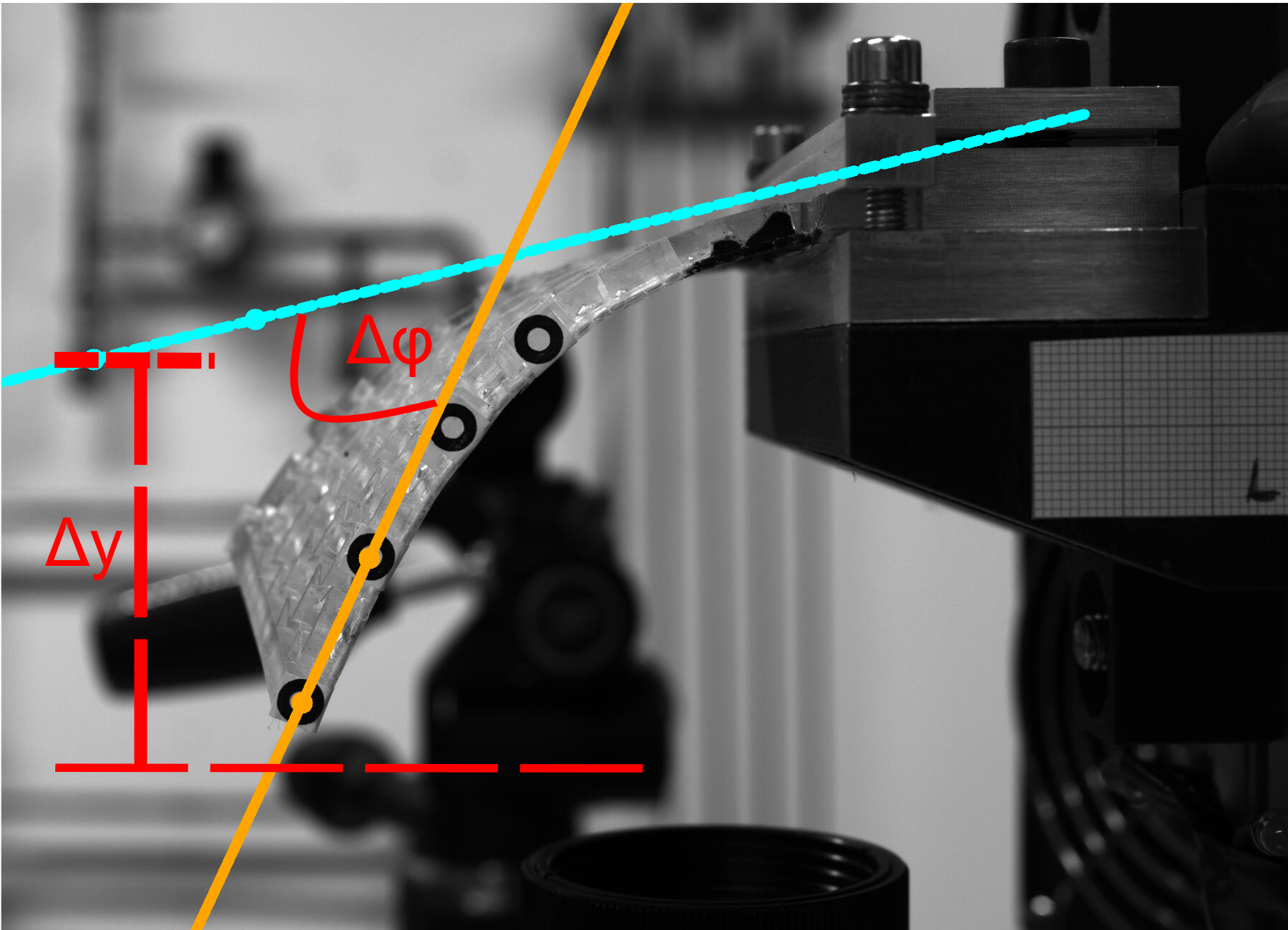}
        \caption{}
        \label{fig:b}
    \end{subfigure}

    \begin{subfigure}[b]{0.48\textwidth}
        \centering
        \includegraphics[width=\linewidth]{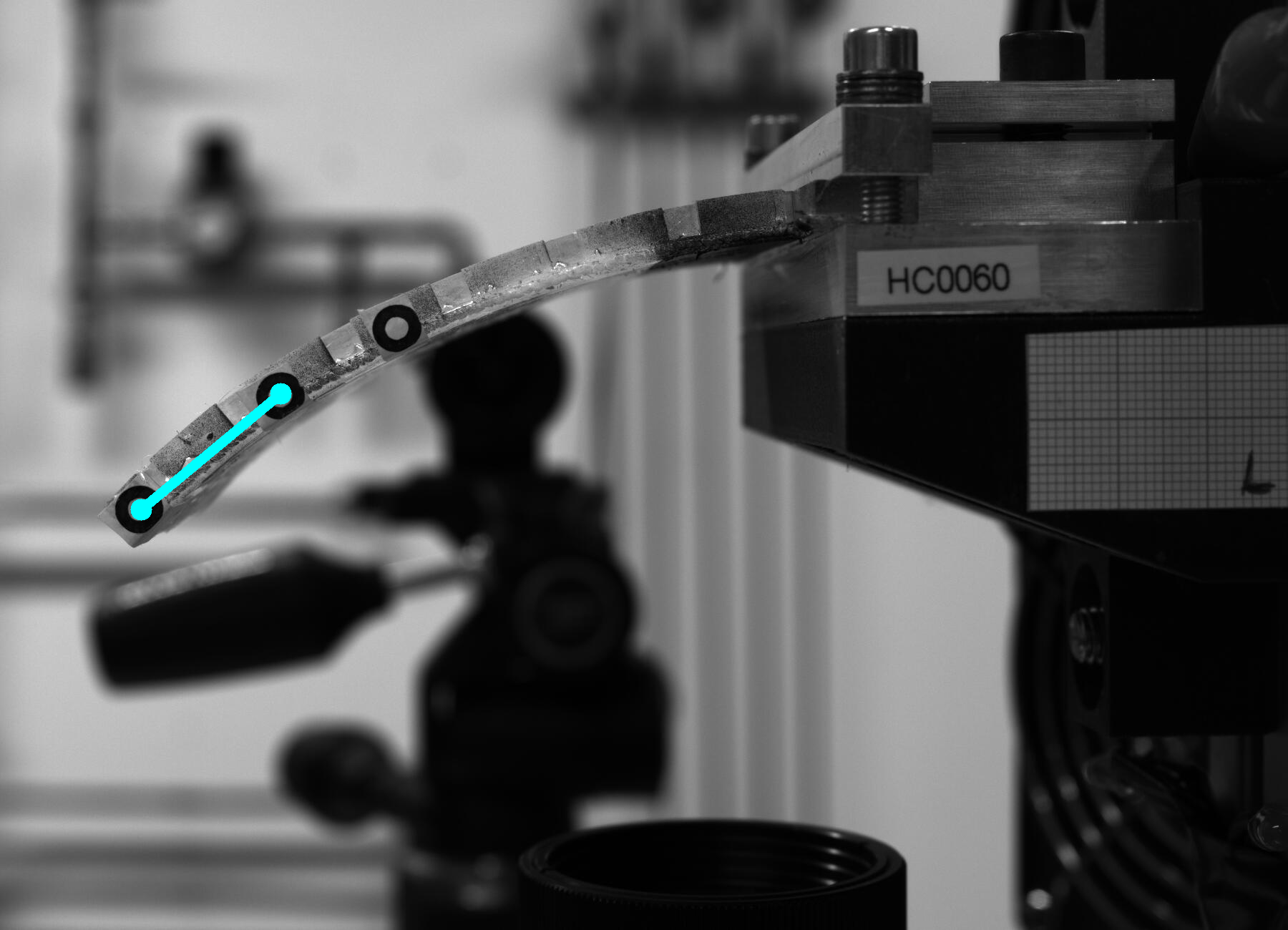}
        \caption{}
        \label{fig:c}
    \end{subfigure}
    \hfill
    \begin{subfigure}[b]{0.48\textwidth}
        \centering
        \includegraphics[width=\linewidth]{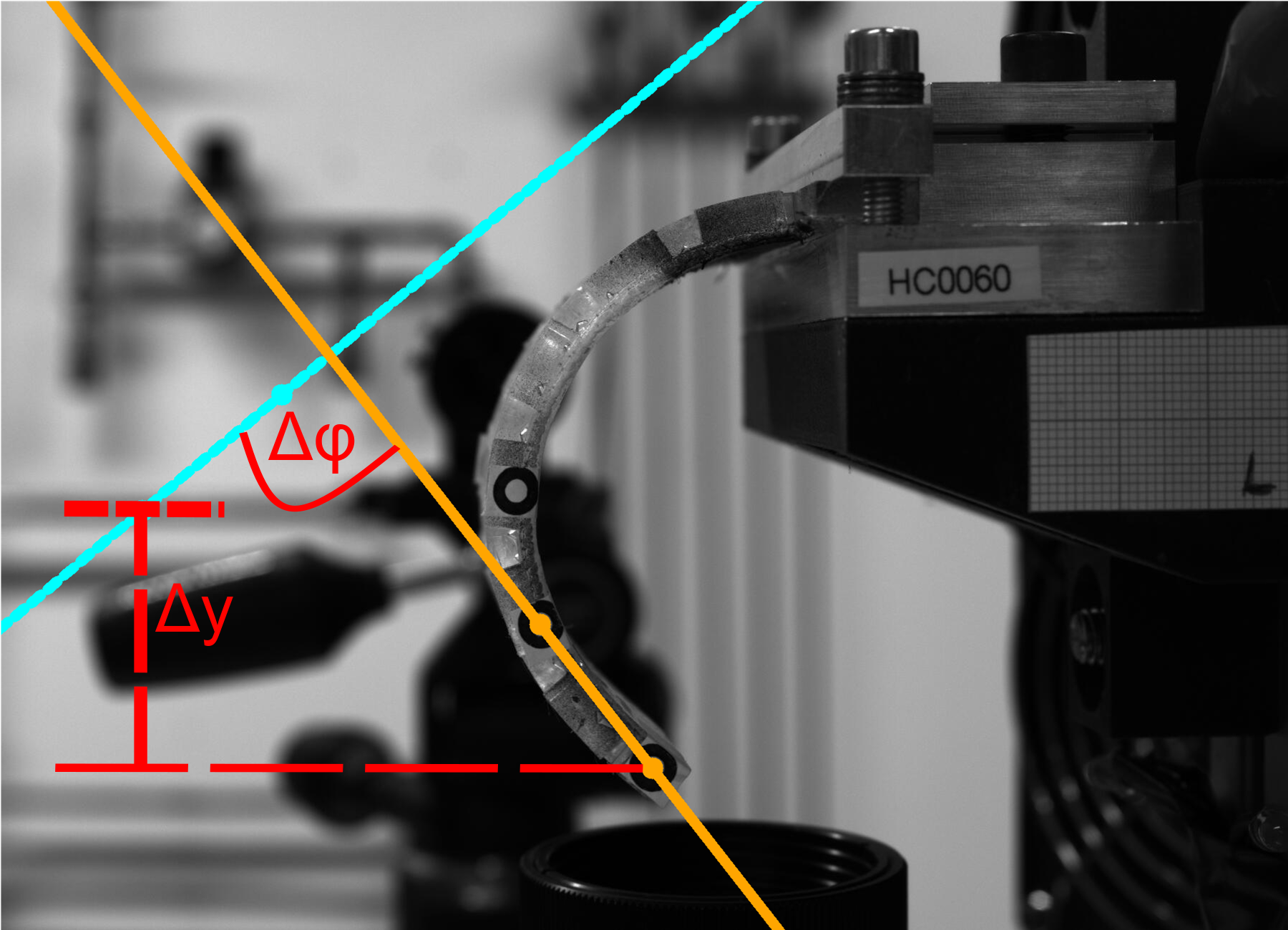}
        \caption{}
        \label{fig:d}
    \end{subfigure}

    \caption{Representative examples of the angle determination based on manually selected points.  
    All images were taken from the respective \textit{leftcam} views.  
    Shown are:  
    a) HC0034 before actuation,  
    b) HC0034 in the fully actuated state,  
    c) HC0060 before actuation, and  
    d) HC0060 in the fully actuated state.  
    For each image, two points were selected and connected by a line segment, and the angle between the initial and actuated orientations was subsequently evaluated}
    \label{fig:angle_examples}
\end{figure}

During thermal activation, all specimen exhibited a pronounced out-of-plane deformation that was clearly observable during the heating and cooling cycles (Fig.\ref{fig:angle_examples}). Upon Joule heating of the embedded SMA wires, the strain mismatch between the active SMA layer and the passive polymer-based toplayer induces a global bending of the hybrid structure. During the subsequent cooling phase, the deformation reverses and the specimens return close to their initial configuration.
While for small deflections of the actuator the vertical displacement of the specimen tip is sufficiently precise, this measure does not fully capture the kinematic behavior of the specimens at larger deformations. In particular, the actuator motion is not confined to a purely vertical trajectory but can involve lateral displacement components and changes in the orientation of the specimen tip. 
To complement the displacement-based evaluation and to gain additional insight into the deformation mode of the specimens, the deflection angle of the specimen tip was measured for one frame at full deflection. 
In the initial frame, two marker points located near the specimen tip were manually selected based on the applied surface markers. A straight line was then defined through these two points, representing the local orientation of the tip region in the non-actuated state. The same two physical marker locations were identified again in the fully actuated frame, and a second straight line was defined accordingly. By construction, this procedure ensures that the orientation change is evaluated consistently at the same region of the specimen before and after actuation. The deflection angle $\Delta\varphi$ was subsequently calculated as the smallest rotation between the two line orientations, corresponding to the change in tip orientation induced by actuation. This angle therefore represents the global rotational component of the out-of-plane deformation between the initial and fully actuated configurations, as illustrated in Fig. \ref{fig:angle_examples}. For comparison, the vertical displacement $\Delta y$ of the specimen tip is also indicated in Fig. \ref{fig:angle_examples} to illustrate the conventional displacement-based measure. Fig. \ref{fig:angle_examples} shows representative examples based on the left camera view for specimens HC0034 and HC0060 to illustrate the angle determination procedure.
Since the deflection angle is determined from manually selected marker points, the measurement accuracy is primarily governed by the positional uncertainty of these points. To obtain a transparent and conservative estimate of the angular uncertainty, a simple geometric error model was applied. The manual selection of each marker point was assumed to be affected by a maximum positional deviation of $\pm 1\,\mathrm{mm}$ in both horizontal and vertical directions. These components were combined into an effective radial position uncertainty
\begin{equation}
\Delta r = \sqrt{(\Delta x)^2 + (\Delta y)^2}.
\end{equation}
Using the spatial calibration of the camera system, this positional uncertainty was converted into the corresponding pixel-based displacement. The deflection angle was evaluated from two straight line segments defined in the undeformed and fully actuated configurations, respectively. Each line segment has a finite length $L$, corresponding to the distance between the two selected marker points and thus defining the effective lever arm of the angle measurement. Under the assumption that the positional uncertainty acts at both ends of the line segment, the resulting angular uncertainty can be approximated by
\begin{equation}
\Delta \varphi \approx 2 \arctan\left(\frac{\Delta r}{L}\right).
\end{equation}
This expression provides an upper bound for the uncertainty of the measured deflection angle and depends explicitly on the selected marker spacing. For each specimen and camera view, the corresponding line length was used to compute an individual uncertainty estimate. These values were subsequently averaged within each specimen configuration to obtain a representative uncertainty level for the aggregated angle data.

The resulting angular uncertainties are small compared to the specimen-to-specimen variability and the observed configuration-dependent differences, confirming that the manual angle evaluation provides sufficient accuracy for comparative analysis.

\begin{figure}[H]
 \centering
        \includegraphics[width=0.9\textwidth]{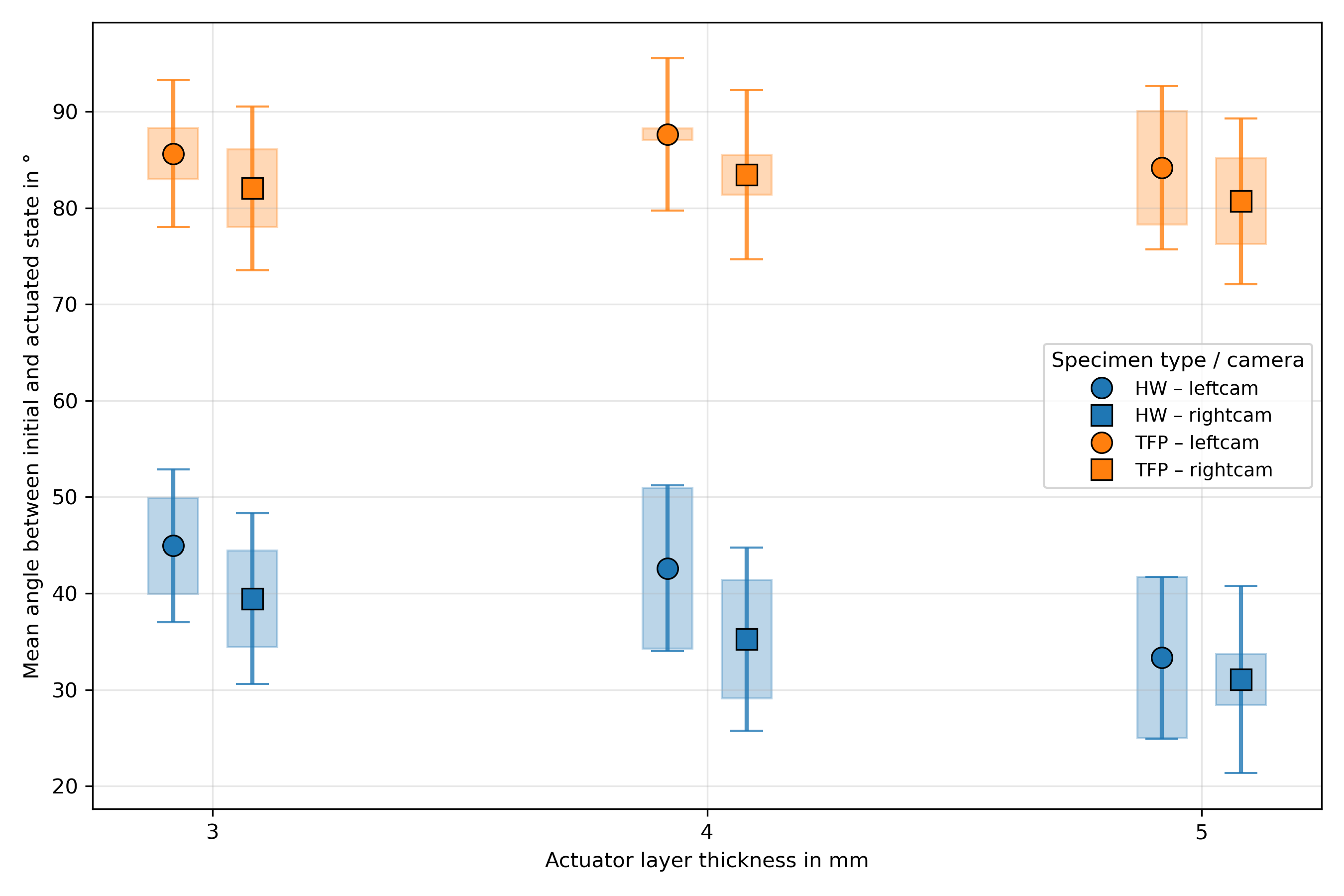}
 \caption{Angle between the non-actuated and fully actuated states for all HW and TFP specimens, grouped by toplayer thickness. Angles were evaluated independently from the left and right camera views. Markers denote the mean deflection angle obtained from three nominally identical specimens per configuration. Shaded rectangles indicate the geometric measurement uncertainty derived from the angle estimation based on a positional marker uncertainty of $\pm 1\,\mathrm{mm}$. Error bars represent the standard deviation across the three specimens within each configuration.}
 \label{angle_vs_thickness_scatter}
\end{figure}

The angle values derived from the manually selected marker pairs were aggregated for all specimens and evaluated as a function of the toplayer thickness. Fig.~\ref{angle_vs_thickness_scatter} summarizes the resulting orientation change between the initial and fully actuated states for both camera views and for all investigated configurations. For each configuration, the central marker denotes the mean deflection angle obtained from the three nominally identical specimens. The shaded rectangles represent the geometric measurement uncertainty of the angle evaluation, derived from the deterministic error model based on a positional uncertainty of ±1 mm for the manually selected marker points and the corresponding marker spacing. The vertical error bars indicate the standard deviation of the deflection angles across the three specimens within each configuration and thus quantify the specimen-to-specimen variability. In all cases, this variability exceeds the geometric measurement uncertainty, indicating that the observed scatter is dominated by physical differences between specimens rather than by limitations of the angle measurement. The HW specimens exhibit markedly lower rotation angles, typically ranging from approximately 20° to 55°. This scatter reflects the sensitivity of the manually fabricated hybrid laminates to small deviations in wire alignment and anchorage conditions. In contrast, the TFP specimens consistently display significantly higher rotation angles, clustering between roughly 75° and 90° for all thicknesses. The combination of larger and highly consistent angle values implies that the TFP architecture promotes a more uniform rotational deformation mode across specimens of the same configuration. Overall, the angle-based evaluation reveals clear and systematic differences between the two manufacturing approaches. While HW specimens show lower angles accompanied by pronounced variability, TFP specimens exhibit a well-defined and tightly grouped rotational response. This characterization provides an independent measure of the deformation mode that complements the displacement-based metrics presented in the subsequent section.

\subsection{Cyclic actuator response and stabilization behavior}
\begin{figure}[H]
    \centering
    \includegraphics[width=0.95\textwidth]{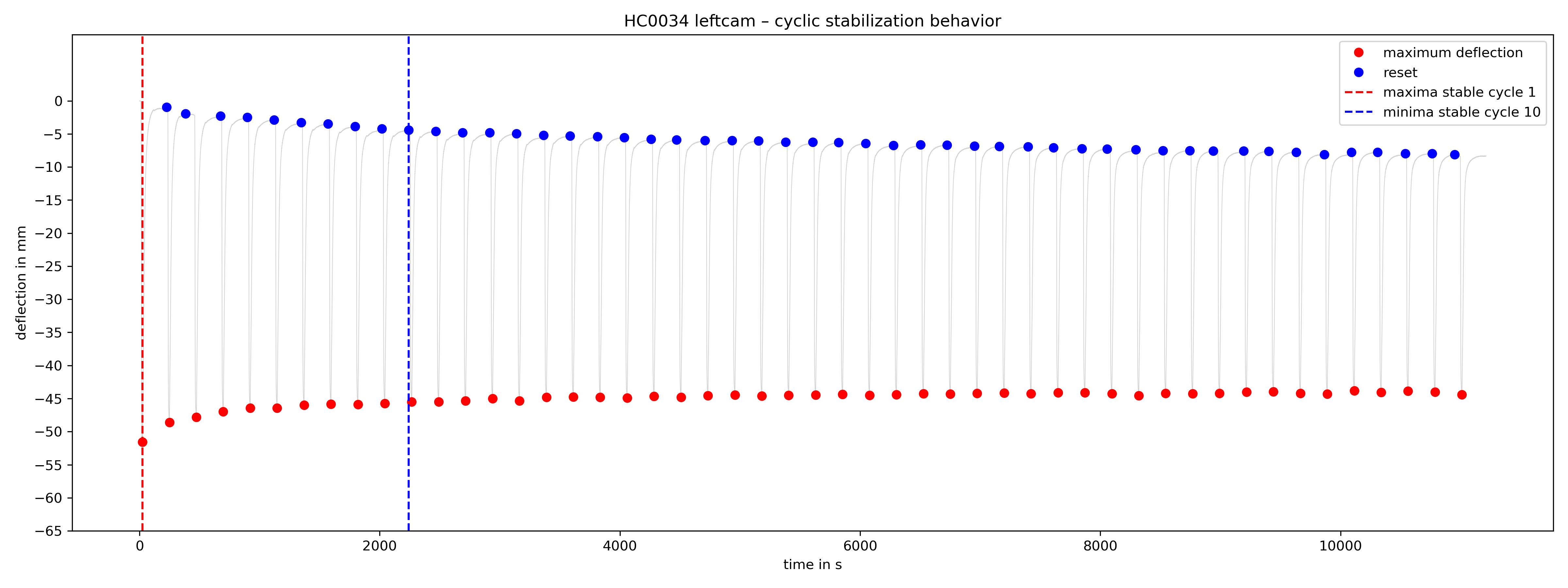}\\
    \includegraphics[width=0.95\textwidth]{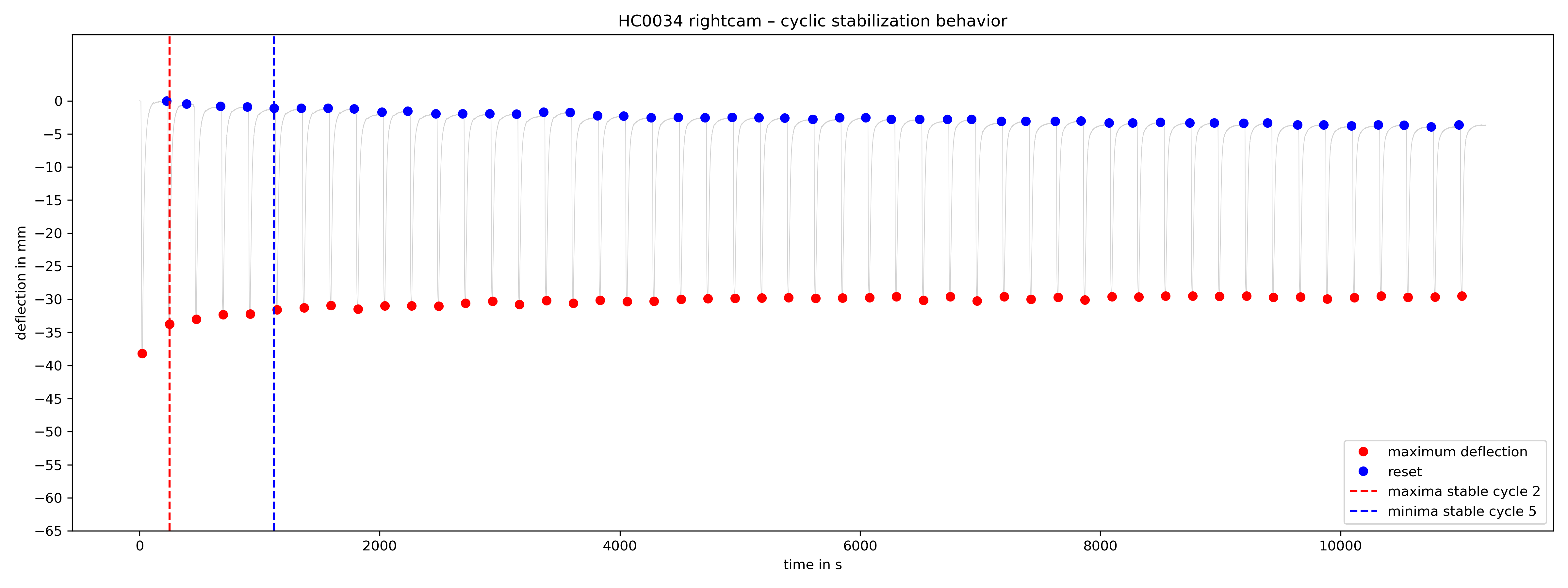}
    \caption{Cyclic stabilization behavior of specimen HC0034 (4 mm HW). Red and blue markers show maximum and reset deflection for each cycle. After an initial running-in phase, both extrema converge toward stable values. Dashed lines indicate the first cycles meeting the ±6 \% stabilization criterion, which define the baseline for subsequent stroke and tilt analysis.}
    \label{fig:HC0034_stabilization}
\end{figure}

To illustrate the cyclic actuator response, Fig.~\ref{fig:HC0034_stabilization} presents the full set of 50 thermal actuation cycles recorded for a representative specimen (HC0034, 4~mm HW configuration) as observed independently by the left and right cameras of the dual-camera system.
The deflection signal, expressed as the relative vertical displacement $\Delta y$, exhibits the characteristic heating and cooling phases associated with the martensite–austenite transformation of the embedded SMA wires. During the first cycles, slightly larger peak displacements and a minor baseline drift are visible, after which the response gradually stabilizes.
This qualitative behavior was observed consistently across all specimens, although the absolute stroke amplitude and transient response depended on specimen thickness and the fiber-integration method. For a clearer interpretation of the stabilization behavior, the maximum deflection (heating peak) and the reset position after cooling were extracted for every cycle. Red markers in Fig.~\ref{fig:HC0034_stabilization} denote the maximum deflection of each cycle, whereas blue markers represent the corresponding reset positions. Peak detection was performed using a prominence threshold of \SI{3.0}{\milli\meter} and a minimum peak distance of 200 samples, ensuring that every actuation cycle contributed exactly one pair of extrema. To quantitatively identify the onset of stable behavior, the extrema sequences were evaluated using a tolerance-band criterion. A sequence was considered \textit{stable} when the variation of the last $n = 4$ consecutive extrema remained within a tolerance band of $\pm 6\%$ around their mean value. Stabilization points were determined independently for the maxima and the reset positions, allowing potential asymmetries between the heating and cooling phases to be captured.

Since the two stabilization points can differ, e.g. maximum deflection becoming stable earlier than the reset position, further analyses consistently reference the later of the two stabilization cycles. This ensures that all subsequent evaluations are based on a fully stabilized cyclic response. From this cycle onward, mean values of the maximum and minimum $\Delta y$ were computed and used to represent the steady-state actuation amplitude of each specimen. This procedure defines a robust and reproducible basis for all subsequent quantitative analyses. The later chapters on stroke and tilt behavior therefore rely exclusively on these stabilized mean deflection values, ensuring that all comparative evaluations reflect the true steady-state actuator performance.

\subsection{Stroke and tilt behavior}
\label{sec:stroke_tilt}
\begin{figure}[h]
  \centering
  \includegraphics[width=0.95\textwidth]{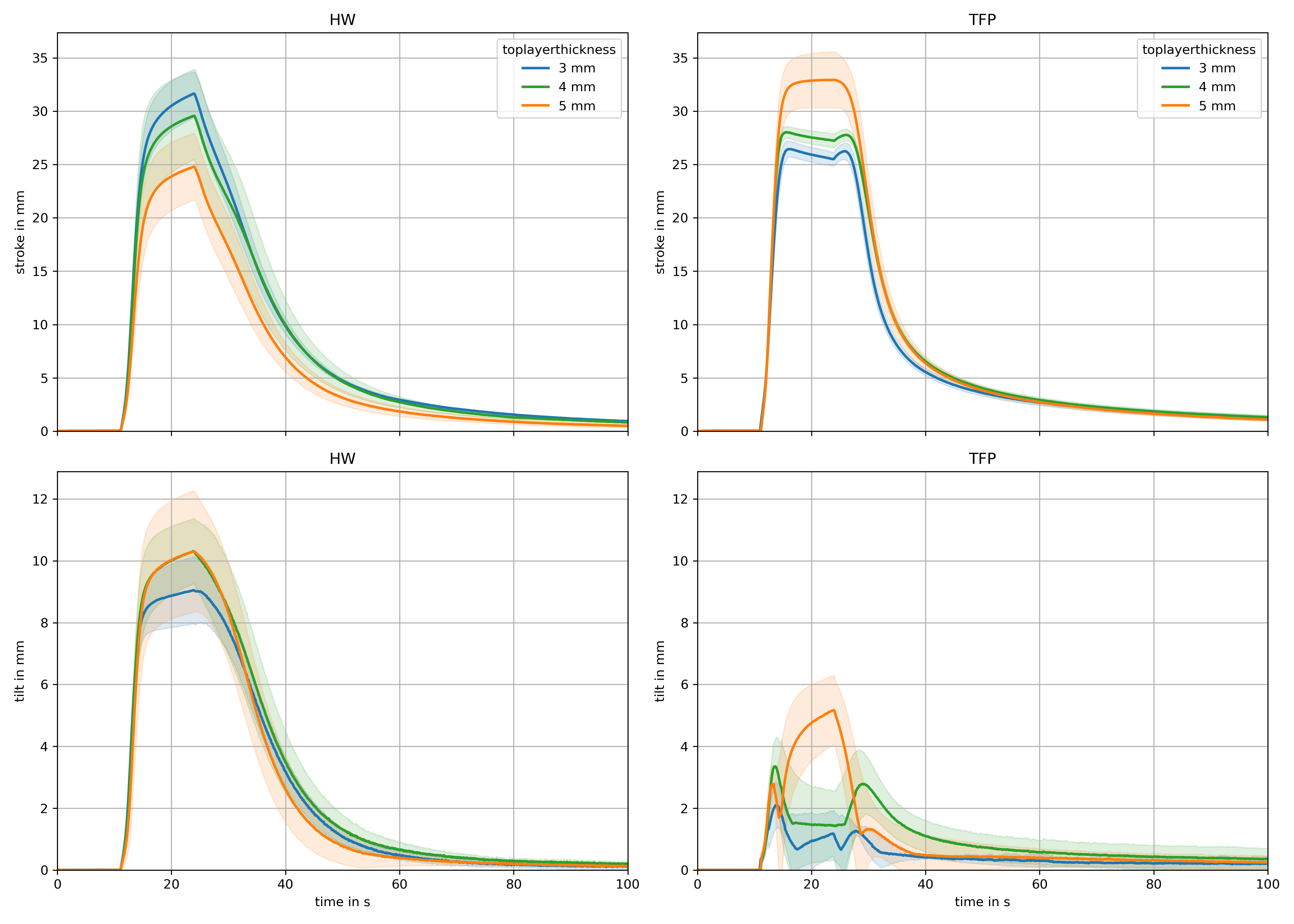}
  \caption{Averaged stroke and tilt responses of SMAHC specimens for HW (left column) and TFP (right column) configurations with identical actuator geometries and different toplayer thicknesses. The upper row shows the stroke evolution, while the lower row displays the corresponding tilt behavior. Solid lines represent the mean response across three nominally identical specimens, and shaded bands indicate the standard deviation. All curves correspond to cycle 20, which lies within the stabilized actuation regime.}
  \label{fig:stroke_tilt_results}
\end{figure}

Although a single-point vertical displacement $\Delta y$ is insufficient to fully describe the geometric deformation mode of the actuator, displacement-based quantities remain highly informative when evaluated in a differential and combined manner. By separating the symmetric and asymmetric components of the vertical edge displacements, additional insight into the actuation performance and deformation stability can be obtained. For this purpose, the deflection signals from both cameras for each specimen were analyzed to determine the global out-of-plane stroke and the lateral tilt component. The stroke quantifies the effective bending amplitude of the actuator, while the tilt captures asymmetric deformation between both edges. The stroke amplitude was defined as the mean vertical displacement at the leading edge, obtained as the average of the absolute deflections measured by the left and right cameras:
\[
\text{stroke} = \frac{\left(|\Delta y_\mathrm{left}| + |\Delta y_\mathrm{right}|\right)}{2}
\]
The tilt amplitude, representing the asymmetric out-of-plane motion of the two edges, was calculated as the absolute difference between both camera signals:
\[
\text{tilt} = |\Delta y_\mathrm{left} - \Delta y_\mathrm{right}|.
\]
This approach allows a clear separation between the global bending deformation and the superimposed asymmetric component, ensuring that both effects are quantitatively comparable across specimens.
Fig. \ref{fig:stroke_tilt_results} summarizes the averaged stroke and tilt responses for HW and TFP specimens with identical actuator geometries and three different toplayer thicknesses. The upper row shows the stroke evolution during heating and subsequent cooling, while the lower row displays the corresponding tilt behavior. Solid lines represent the mean response across three nominally identical specimens, and the shaded bands indicate the standard deviation. For HW specimens, the stroke response exhibits a pronounced dependence on the toplayer thickness. Thinner configurations reach higher peak stroke amplitudes, while increasing thickness leads to a systematic reduction of the achievable bending. In addition, the HW configurations show a comparatively slow and gradual decay during the cooling phase. The corresponding tilt responses reveal substantial lateral asymmetry during actuation, with peak tilt values that are comparable in magnitude to the stroke-induced displacement at the edges. In contrast, TFP specimens exhibit a markedly different behavior. The stroke amplitudes are consistently higher across all thicknesses and show a pronounced plateau during the heating phase. The plateau does not indicate a complete saturation of the deformation. Instead, this behavior reflects a limitation of the displacement-based measurement approach. Beyond a certain deformation state, the actuator motion increasingly deviates from the vertical direction, and additional deformation occurs predominantly in the lateral (x-) direction. Since the stroke evaluation is based solely on the vertical displacement component, this lateral contribution is not captured, resulting in an apparent plateau in the stroke signal. This observation is consistent with the angle-based analysis presented earlier, which revealed a significant rotational component of the deformation for TFP specimens. The plateau therefore highlights the geometric transition of the deformation mode rather than a limitation of the actuator itself. Overall, the combined stroke and tilt analysis reveals that the TFP architecture not only increases the effective actuation amplitude but also promotes a more symmetric and temporally stable deformation behavior. The tilt response of TFP specimens remains substantially lower than that of HW specimens throughout the entire actuation cycle, demonstrating a strongly reduced lateral asymmetry. Furthermore, the cooling phase is significantly accelerated, leading to a steeper decay of the stroke signal. 

\begin{figure}[htbp]
  \centering
  \includegraphics[width=0.8\textwidth]{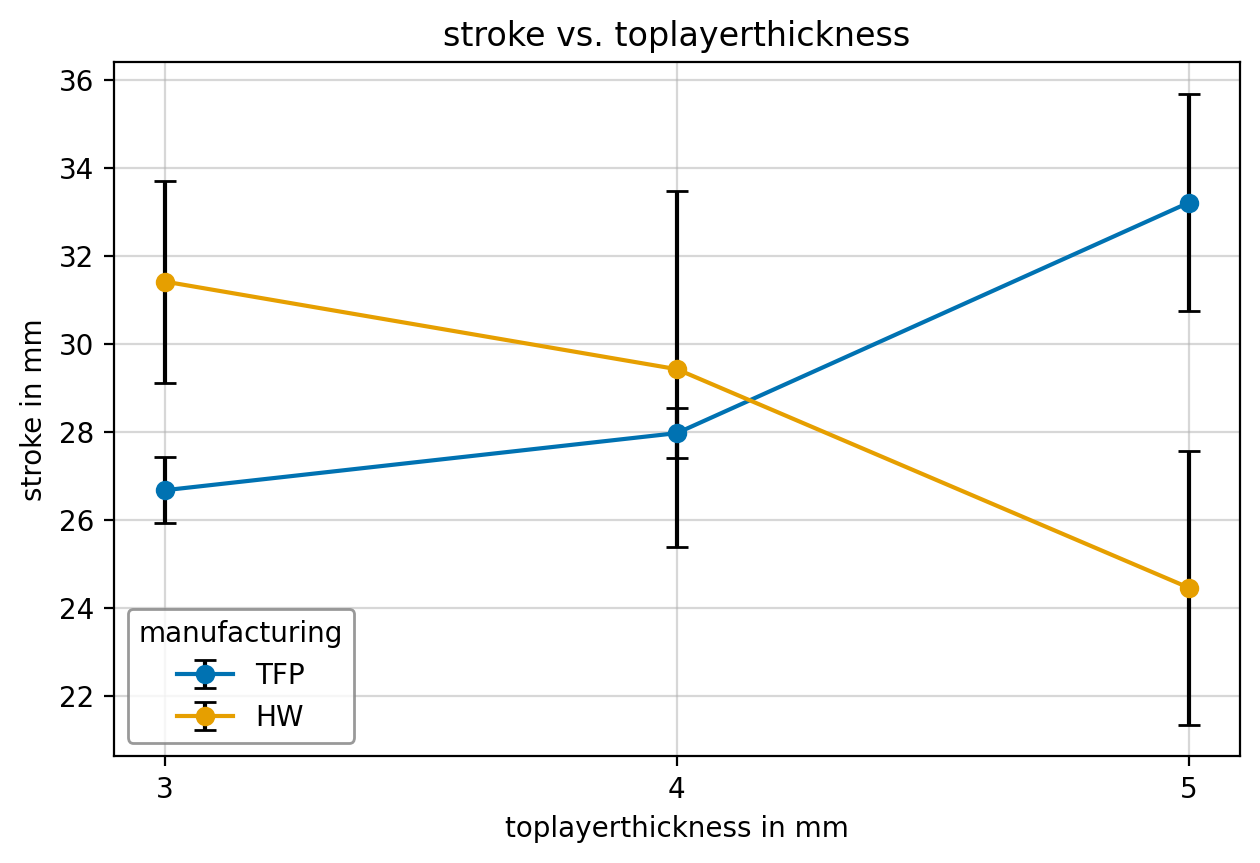}
  \caption{Mean stabilized stroke amplitude as a function of toplayer thickness for HW and TFP SMAHC specimens. Data points represent mean values obtained from three nominally identical specimens per configuration. Error bars indicate one standard deviation.}
  \label{fig:stroke_vs_thickness}
\end{figure}

Figure~\ref{fig:stroke_vs_thickness} summarizes the mean stroke amplitude as a function of the toplayer thickness for both manufacturing approaches. 
For the HW specimens, the stroke amplitude decreases slightly with increasing toplayer thickness, indicating the expected stiffening effect of the thicker polymer face sheets. 
In contrast, the TFP-based laminates exhibit a moderate increase in stroke amplitude with thickness. Despite the similar overall geometry, the TFP specimens consistently show a smaller standard deviation, suggesting a higher reproducibility of the fiber placement and bonding quality.

\begin{figure}[htbp]
  \centering
  \includegraphics[width=0.8\textwidth]{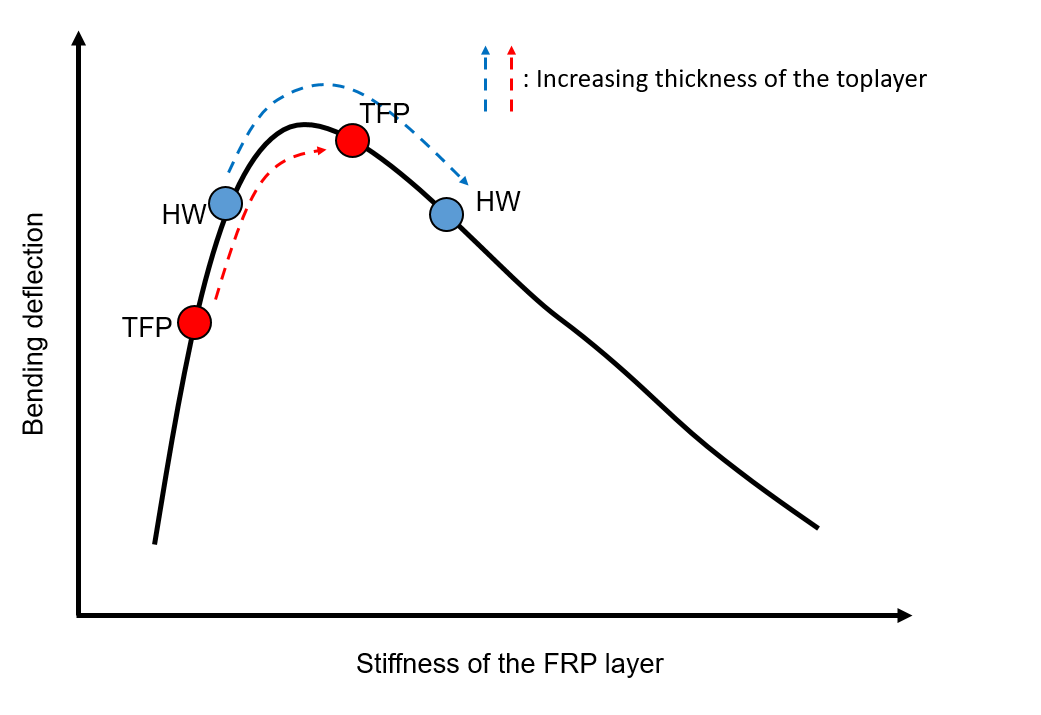}
  \caption{Qualitative dependence of bending deflection on the stiffness of the passive FRP layer (toplayer). TFP and HW specimens occupy different regions of the curve due to their distinct compressive stiffness: TFP (unidirectional) lies on the compliant side, while HW (multidirectional) lies closer to the over-stiff regime. Increasing the toplayer thickness shifts both configurations along the curve, as indicated by the dashed arrows.}
  \label{fig:schematic_bending_VS_stiffness}
\end{figure}

The opposite thickness-dependence of the stroke for HW and TFP specimens can be explained by the qualitative relation between bending deflection and the stiffness of the passive FRP layer (toplayer). Fig. \ref{fig:schematic_bending_VS_stiffness} provides a schematic representation of this behavior. The bending curvature generated by an active layer first increases with the stiffness of the passive substrate, because a stiffer backing suppresses local compression and shear deformation and therefore converts a larger fraction of the recoverable SMA strain into global bending. Beyond an intermediate optimum, however, further stiffening of the passive layer mainly increases the bending stiffness of the composite without significantly improving the load transfer, so that the achievable curvature, and thus the stroke decreases again \cite{gurka_physics_2019}. The HW and TFP architectures occupy different regions on this qualitative curve, as indicated by the colored markers in Fig.\ref{fig:schematic_bending_VS_stiffness}. The HW specimens employ a multidirectional glass-fiber lay-up (see section  \ref{sec:method and materials}), which provides comparatively high compressive stiffness of the passive layer already at the smallest toplayer thickness. Consequently, the initial HW datapoint lies on the rising part of the curve but relatively close to the optimum. Increasing the HW toplayer thickness therefore shifts the structure further into the “over-stiff” regime of the diagram (blue arrow), in which the neutral axis moves toward the laminate mid-plane and the global bending stiffness dominates. As a result, the additional thickness predominantly penalizes curvature, and the measured stroke monotonically decreases with thickness.
In contrast, the TFP actuatorlayer is almost purely unidirectional. In through-thickness compression it is therefore much more compliant, even though the rovings provide efficient tensile and bending reinforcement along the actuator length. The initial TFP datapoint consequently lies far on the left side of the curve, in a region where the passive FRP layer (toplayer) is still too soft to effectively transfer the SMA recovery strain into global bending. Increasing the toplayer thickness in the TFP architecture stiffens the passive backing and increases the distance between the SMA wires and the neutral axis. This moves the system upward and rightward along the curve (red arrow), toward its optimum region. In this regime, the increase in transformation efficiency outweighs the rise in composite bending stiffness, which explains the moderate increase in stroke amplitude.

\begin{figure}[htbp]
  \centering
  \includegraphics[width=0.8\textwidth]{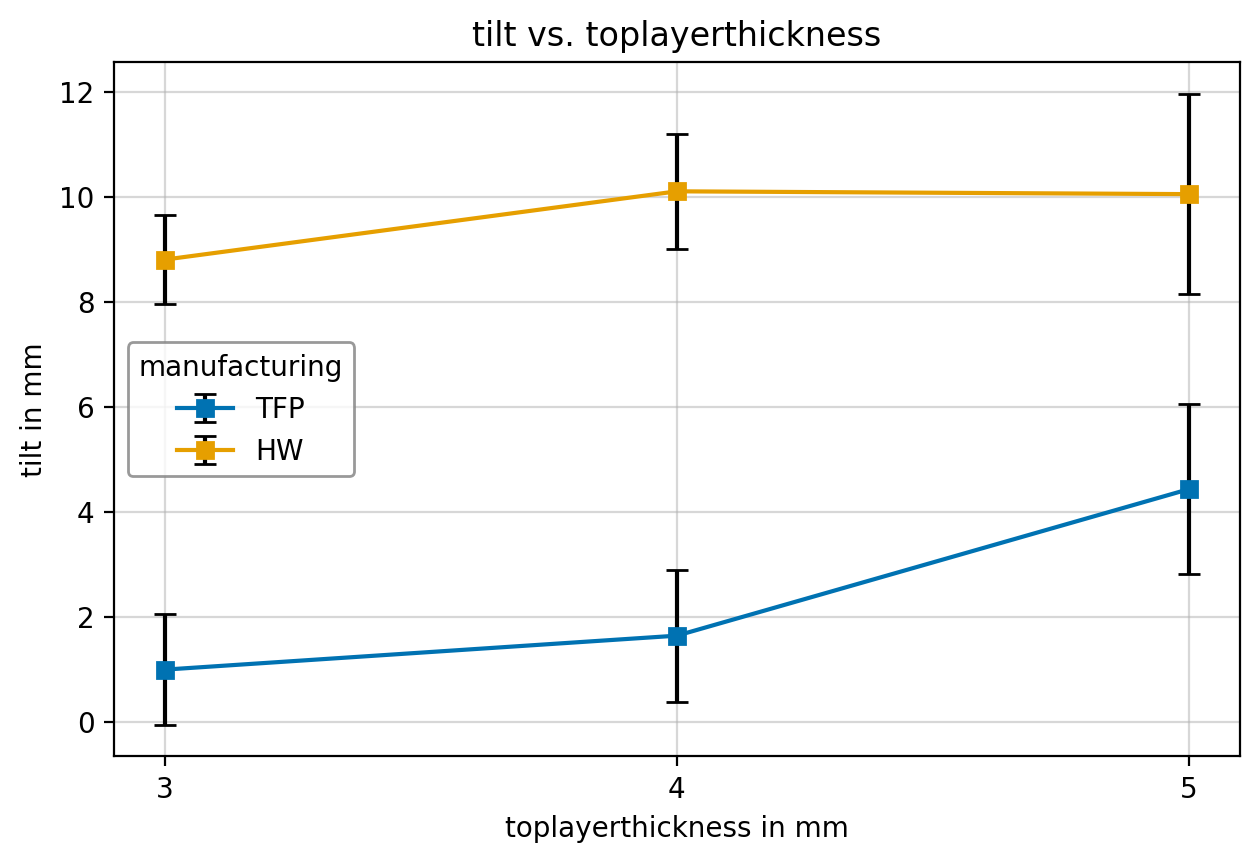}
  \caption{Mean stabilized tilt amplitude as a function of toplayer thickness for HW and TFP SMAHC specimens. Data points represent mean values obtained from three nominally identical specimens per configuration. Error bars indicate one standard deviation.}
  \label{fig:tilt_vs_thickness}
\end{figure}

The corresponding tilt amplitudes, shown in Fig.~\ref{fig:tilt_vs_thickness}, reveal distinct differences between the integration methods. 
HW specimens show substantially larger and less consistent tilt amplitudes compared to TFP samples. 
This behavior suggests a more asymmetric actuation response, likely caused by non-uniform SMA wire embedding and local fiber misalignment inherent to the manual hand-weaving process. 
In contrast, the TFP samples exhibit a systematic, nearly linear increase in tilt with thickness, while maintaining smaller absolute tilt magnitudes and standard deviations. 
The improved symmetry and reduced scatter in the TFP specimens confirm the advantages of automated wire placement for achieving controlled actuation.

\begin{figure}[htbp]
  \centering
  \includegraphics[width=0.8\textwidth]{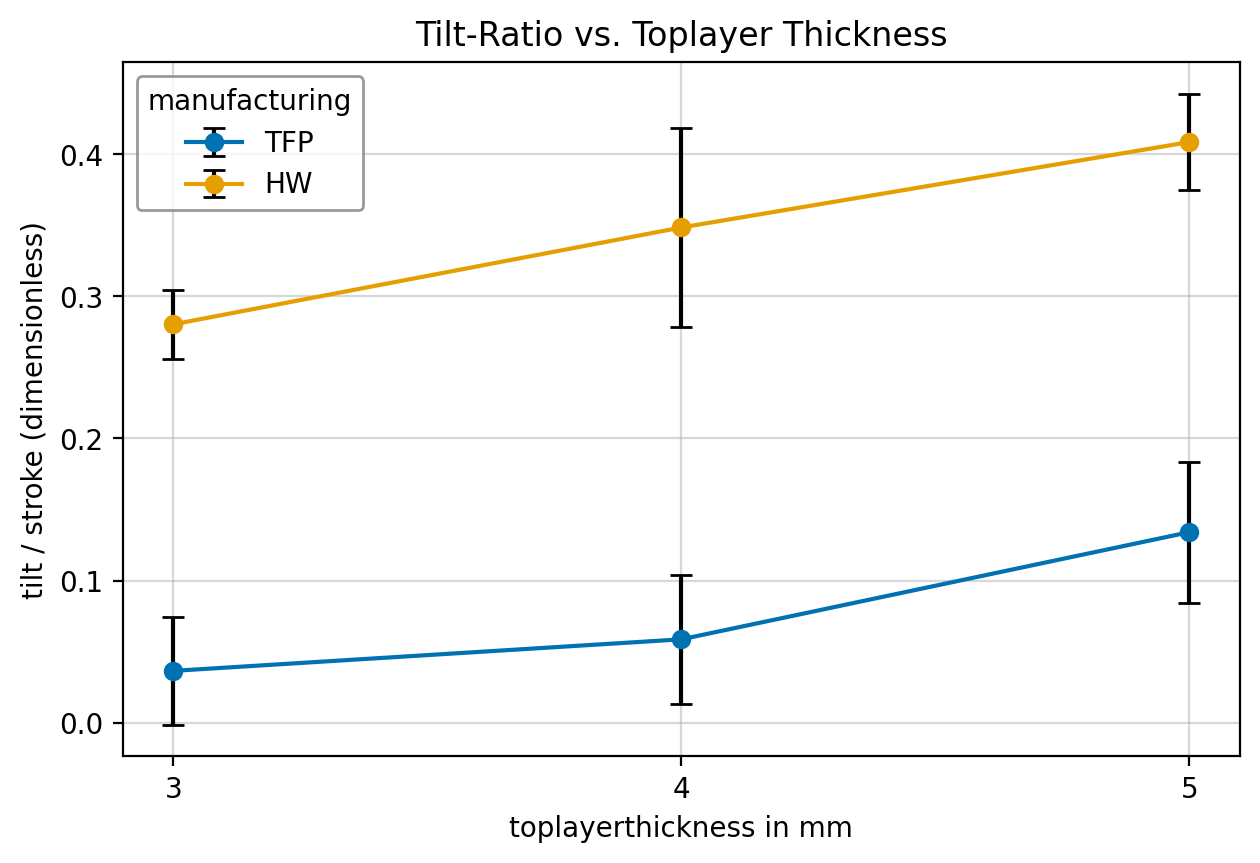}
  \caption{Tilt-to-stroke ratio as a function of toplayer thickness for HW and TFP SMAHC specimens. The ratio provides a normalized measure of asymmetric deformation relative to the total bending amplitude. Data points represent mean values, with error bars indicating one standard deviation.}
  \label{fig:tilt_ratio_vs_thickness}
\end{figure}

To further quantify the relationship between symmetric and asymmetric actuation, the ratio of tilt to stroke (``tilt ratio'') was computed for each specimen, as illustrated in Fig.~\ref{fig:tilt_ratio_vs_thickness}. 
This dimensionless quantity expresses the degree of asymmetry normalized by the total bending amplitude. 
The results show that HW structures exhibit tilt ratios roughly three to four times higher than those of the TFP samples, particularly for thicker face sheets. 
The trend towards increasing tilt ratio with thickness in both configurations indicates that stiffer toplayers amplify minor embedding imperfections, which in turn translate into a more pronounced asymmetric deformation.

\begin{figure}[H]
  \centering
  \includegraphics[width=0.8\textwidth]{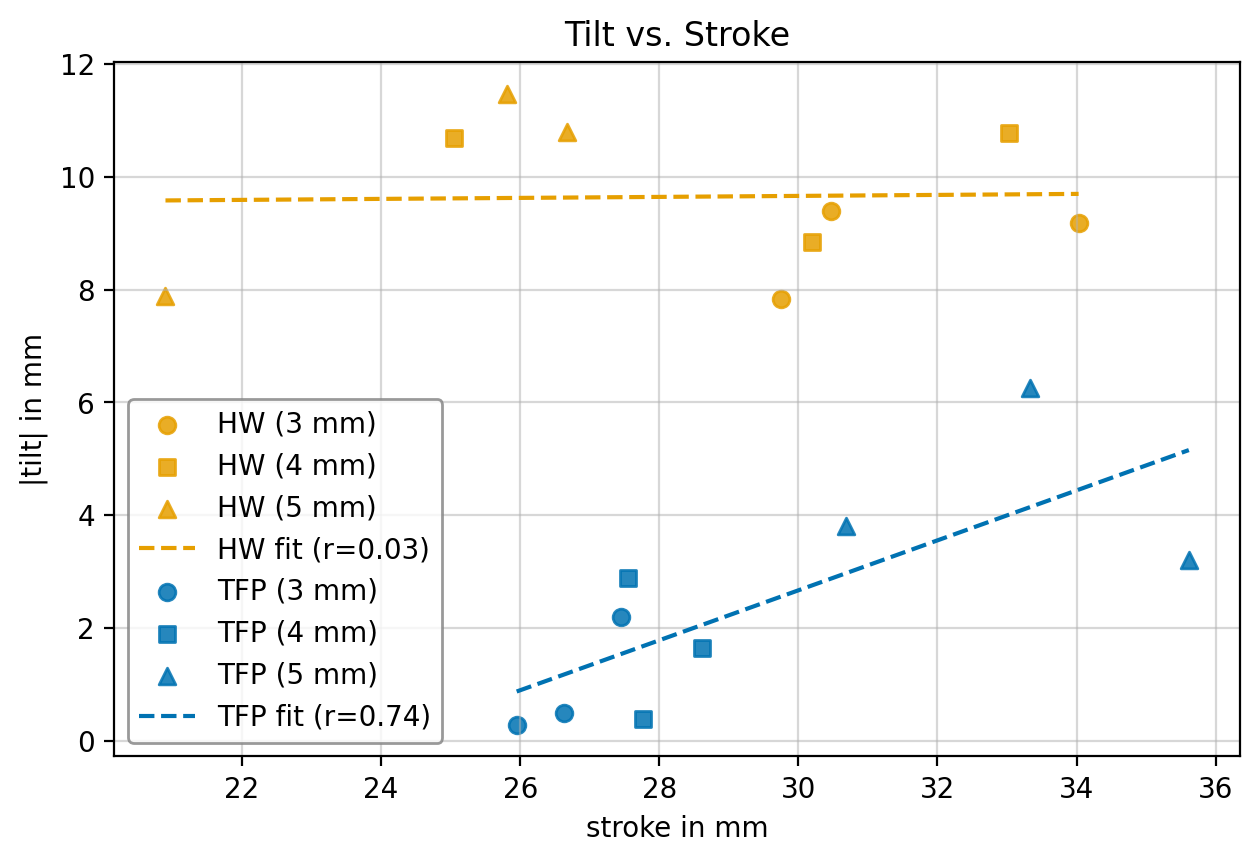}
  \caption{Correlation between stabilized stroke amplitude and corresponding tilt amplitude for all investigated SMAHC specimens. Each data point represents the mean stabilized response of one specimen. HW and TFP configurations are shown for all toplayer thicknesses. Dashed lines denote linear regression fits, illustrating the strong stroke–tilt coupling observed for TFP specimens and the absence of a systematic correlation for HW specimens.}
  \label{fig:tilt_vs_stroke}
\end{figure}

\begin{figure}[H]
    \centering
    \includegraphics[width=0.95\linewidth]{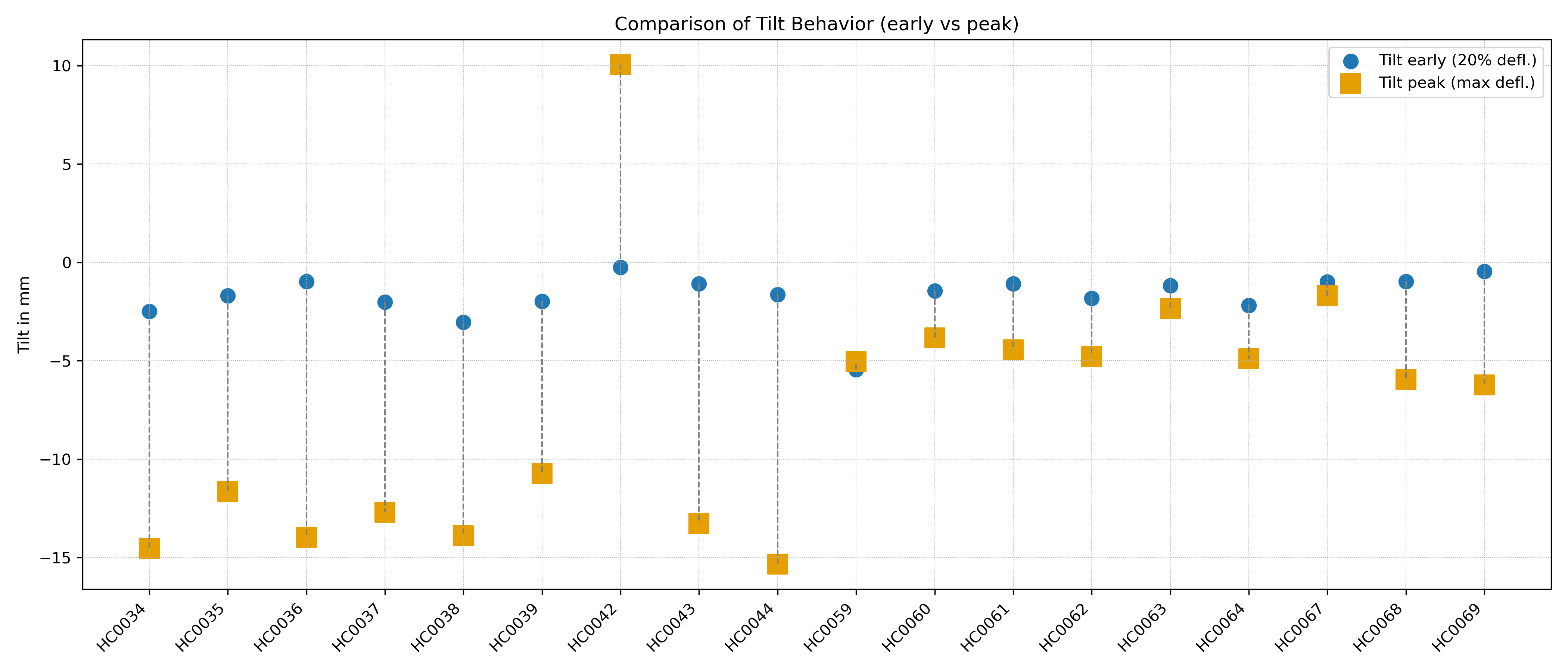}
    \caption{Comparison of early-state tilt (20 \% of maximum deflection) and peak tilt amplitude for all investigated SMAHC specimens. Results are shown for HW and TFP configurations for all toplayer thicknesses. While early-state tilt remains small and comparable for both manufacturing routes, pronounced differences emerge at peak actuation.}
    \label{fig:tilt_early_peak}
\end{figure}

A Pearson correlation analysis between stroke and tilt amplitudes is shown in Fig.~\ref{fig:tilt_vs_stroke}. The TFP specimens exhibit a clear positive correlation ($r = 0.74$), indicating that larger stroke amplitudes are associated with proportionally higher tilt components. This behaviour suggests a systematic geometric coupling between axial actuation and asymmetric bending. In contrast, the HW specimens display almost no correlation ($r = 0.03$), resulting in a nearly random distribution of peak tilt values. This decoupling between stroke and tilt in the HW structures points to process-induced variability as the dominant source of the asymmetric response rather than an intrinsic geometry-driven coupling mechanism.
To determine whether this asymmetry is already present at the beginning of the actuation cycle or develops only towards the maximum stroke, an additional analysis of the tilt at an early deformation state was performed, as presented in Fig.~\ref{fig:tilt_early_peak}. For this purpose, the tilt amplitudes at 20\,\% of the maximum deflection (``early state'') were compared with the corresponding peak tilt values for all specimens. The early-state tilt values of both manufacturing routes are remarkably similar and remain comparatively small in magnitude. The initial hypothesis that HW specimens exhibit a distinct tilt from the onset of the cycle is therefore not supported. Instead, the characteristic differences between HW and TFP specimens emerge only during the progression from early deformation to peak stroke.
In the TFP specimens, the tilt increase from the early state to the maximum stroke scales consistently with the stroke magnitude, which is fully consistent with the strong stroke--tilt correlation observed in Fig.~\ref{fig:tilt_vs_stroke}. This behavior reflects a systematic and predictable geometric coupling introduced by the TFP architecture. In contrast, the HW specimens show a highly scattered increase in tilt with no discernible relation to the stroke amplitude. The additional tilt generated during the actuation cycle is dominated by local imperfections and uncontrolled asymmetries inherent to the manual integration process.

Overall, the results demonstrate that the TFP integration yields a more homogeneous and predictable actuation behaviour, whereas the HW approach, while capable of achieving larger absolute strokes in some cases, suffers from increased asymmetry and reduced reproducibility. These findings confirm that the automated TFP process offers superior control over both the magnitude and the symmetry of the SMA-induced deformation in re-entrant SMAHC architectures.

\section{Conclusion and Outlook}
\label{sec:conclusion}
This work demonstrated a fully additive and reproducible manufacturing approach for the fabrication and functional evaluation of shape-memory alloy hybrid composites. By combining stereolithography with embedded SMA-reinforced glass-fiber layers, the presented method enables the realization of SMAHC structures with tailored toplayer geometries produced in a single, fully integrated fabrication process. In contrast to previous approaches, which relied on sequential or multi-step processing routes, the method introduced here allows geometrically complex, fiber-reinforced, and functionally graded composite structures to be produced seamlessly within one additive manufacturing workflow.

The experiments confirm that the resulting specimens exhibit consistent, reversible, and thermally induced shape adaptation, with stable actuation performance achieved after only a short stabilization phase. The influence of toplayer stiffness on stroke amplitude and symmetry could be quantified precisely using synchronized dual-camera measurements, further validating the robustness and reproducibility of the fabrication process.

A central advantage of the approach lies in its unprecedented design freedom:
the additively manufactured toplayer can be programmed with customized stiffness profiles through controlled geometry, wall thickness, and local reinforcement architecture. This enables not only global stiffness adjustments but also spatially localized stiffness variations, creating regions with specific mechanical functions within the same structural component. Beyond linear stiffness tuning, the additive process also permits the fabrication of architected materials and metamaterials, including auxetic or lattice-based structures, which can shape the deformation field and bending response in ways that are not accessible through conventional composite manufacturing techniques.

This high degree of integrability marks an important step toward monolithic, multifunctional SMAHCs in which actuator, reinforcement, and substrate are fabricated in a single consolidated manufacturing operation, eliminating alignment errors, interfacial weaknesses, or process-induced variability commonly observed in multi-stage processes.

Future work should exploit this capability to realize toplayers with spatially variable thicknesses, internal mesostructures, or segmentation strategies to achieve advanced actuation patterns and programmable deformation fields. Moreover, long-term cyclic stability and thermo-mechanical fatigue need to be assessed to characterize durability under repeated activation. Improving thermal management, e.g. through structured aluminum layers as proposed in \cite{baytekin-gerngross_making_2016}—represents another promising route to enhance cooling efficiency and actuation frequency. Finally, scaling the process through larger stamp-tools, higher-precision alignment fixtures, and multi-material deposition technologies will enable the development of larger, more complex, and more functionally integrated SMAHC demonstrators.

Overall, the combination of SLA-based additive manufacturing and embedded SMA reinforcement establishes a powerful platform for the development of next-generation morphing, thermally adaptive, and smart composite systems, offering a level of geometric and functional customization that was previously unattainable.
%
%

\ack{Sample text inserted for demonstration.}

\funding{This work was funded by Deutsche Forschungsgesellschaft DFG (German Research Foundation) under grant number 404403710.}

\roles{Manuel Kunzler: \\
Sascha Bruk:  \\
Max Kaiser:  \\
Martin Gurka: \\}



\printbibliography

\end{document}